\pdfoutput=1
\documentclass[11pt,letterpaper]{article}
\usepackage[margin=1in]{geometry}
\usepackage{amsmath,amssymb,amsthm,mathtools}
\usepackage{enumitem}
\usepackage{array}
\usepackage{xcolor}
\usepackage{tikz}
\usetikzlibrary{decorations.pathreplacing,decorations.pathmorphing,arrows.meta,positioning,calc,patterns}
\usepackage[colorlinks=true,linkcolor=blue!50!black,citecolor=blue!50!black,urlcolor=blue!50!black]{hyperref}

\newtheorem{theorem}{Theorem}[section]
\newtheorem{lemma}[theorem]{Lemma}
\newtheorem{corollary}[theorem]{Corollary}
\newtheorem{proposition}[theorem]{Proposition}

\newtheorem{claim}[theorem]{Claim}
\theoremstyle{definition}
\newtheorem{definition}[theorem]{Definition}

\DeclareMathOperator{\col}{col}
\DeclareMathOperator{\id}{id}
\DeclareMathOperator{\mass}{mass}

\newcommand{\emitf}{\mathsf{emit}}

\tikzset{
  cb/.style={circle,draw,fill=black,inner sep=1.6pt},
  cw/.style={circle,draw,fill=white,inner sep=1.6pt},
  cg/.style={circle,draw,fill=gray!55,inner sep=1.6pt},
  cl/.style={circle,draw,fill=blue!35,inner sep=1.6pt},
}
\colorlet{cR}{red!75!black}
\colorlet{cB}{blue!75!black}
\colorlet{cG}{green!50!black}
\tikzset{
  vtx/.style={circle,draw=#1,fill=#1!18,line width=0.9pt,inner sep=0pt,minimum size=0.46cm,font=\scriptsize},
  bar/.style={line width=2.6pt,#1},
  wbar/.style={line width=2.6pt,#1,dash pattern=on 3pt off 1.6pt},
  box/.style={draw,rounded corners=2pt,align=center,font=\scriptsize,inner sep=3pt,text width=#1,execute at begin node={\hyphenpenalty10000\relax}},
}
\DeclareMathOperator{\Base}{Base}

\newcommand{\cL}{\mathcal L}

\title{Adjacency Labeling Schemes of Size $\log n+O(\log\log^\ast n)$\\ for Graphs of Bounded Treewidth}
\author{Noy Rotbart\\ {\small Churney, Copenhagen, Denmark}\\ {\small\texttt{noy@churney.io}}}
\date{}

\begin{document}
\maketitle

\begin{abstract}
Gavoille and Labourel (ESA 2007) gave an adjacency labeling scheme of size $\log n+O(k\log\log n)$ for graphs of treewidth at most $k$, and asked whether $\log n+O(k)$ bits suffice. We give a labeling scheme of size $\log n+O(\log\log^\ast n)$ for every fixed $k$, the first improvement of the second-order term for any $k\ge 2$. Equivalently, there is an induced-universal graph with $n(\log^\ast n)^{O_k(1)}$ vertices for this family.

We work with laminar families of intervals, in which a vertex finds each earlier neighbor as the owner of an interval of a given nesting depth around it. As long as every subtree occupies consecutive positions, the nesting depth must be logarithmic, and $\log\log n$ bits per neighbor follow. We give up contiguity. Heavy subtrees are moved out, which brings the depth down to $O(\log^\ast n)$; the intervals broken by the moves are repaired with $2^{O(k^3)}$ additional intervals per vertex; and the bits needed to find these intervals are paid for by the moved subtrees.
\end{abstract}

\section{Introduction}

A labeling scheme is a method of distributing the information about the structure of a graph among its vertices, by assigning a short binary label to every vertex, such that a fixed function on pairs of vertices can be computed from their labels alone. In an \emph{adjacency labeling scheme} for a family $\mathcal F$ of graphs, the function is adjacency: given the labels of two vertices $u,v$ of a graph $G\in\mathcal F$, a decoder decides whether $uv\in E(G)$. The \emph{size} of the scheme is the maximum label size over the $n$-vertex graphs of $\mathcal F$. Hereafter, we refer to adjacency labeling schemes simply as labeling schemes. Kannan, Naor and Rudich~\cite{KNR92} observed that a labeling scheme of size $f(n)$ for $\mathcal F$ is equivalent to an induced-universal graph\footnote{A graph $U$ is induced-universal for a family $\mathcal F$ if every graph in $\mathcal F$ on $n$ vertices appears as an induced subgraph of $U$.} with $2^{f(n)}$ vertices. In light of this equivalence every single bit counts: a labeling scheme of size $\log n+O(1)$ corresponds to an induced-universal graph of linear size.

\subsection{Previous Work}

Trees admit a labeling scheme of size $2\log n$, in which a vertex stores its identifier and that of its parent~\cite{KNR92}. Determining the optimal size for trees was a question raised repeatedly for more than two decades, until Alstrup, Dahlgaard and Knudsen~\cite{ADK17} presented a labeling scheme of size $\log n+O(1)$, which is optimal up to an additive constant.

For graphs of treewidth at most $k$, Gavoille and Labourel~\cite{GL07} presented a labeling scheme of size $\log n+O(k\log\log(n/k))$ and proved a lower bound of $\log n+\Omega(k)$. They asked whether a labeling scheme of size $\log n+O(k)$ exists, noting that for $k=1$ this includes the question for trees, which was open at the time. To the best of our knowledge, the second-order term $O(k\log\log n)$ has not been improved for any $k\ge 2$.

The labeling scheme of~\cite{GL07} is also a building block of the labeling schemes for planar graphs, of size $(4/3+o(1))\log n$~\cite{BGP20} and $(1+o(1))\log n$~\cite{DEGJMM21}. Dujmovi\'c et al.~\cite{DEGJMM21} label the subgraphs of $H\boxtimes P$, where $H$ has bounded treewidth and $P$ is a path, with $\log n+O(\sqrt{\log n\log\log n})$ bits, and list the right lower-order term as an open problem; graphs of bounded treewidth are the case in which $P$ has a single vertex. Labeling schemes of size $(1+o(1))\log n$ are now known for every proper minor-closed family~\cite{DGJMMW26}.

\subsection{Our Contribution}

Our main contribution is a labeling scheme of size $\log n+O(\log\log^\ast n)$ for graphs of bounded treewidth.\footnote{$\log^\ast n$ is the number of times $\log$ should be iterated, starting from $n$, before the result is at most $1$.} To the best of our knowledge, this is the first improvement of the second-order term of~\cite{GL07} for any $k\ge 2$. We summarize this result in the following theorem.

\begin{theorem}\label{thm:main}
For every $k\ge 1$, the family of graphs of treewidth at most $k$ admits a labeling scheme of size $\log n+2^{O(k^3)}(\log\log^\ast n+1)$. In particular, the size is $\log n+O(\log\log^\ast n)$ for every fixed $k$.
\end{theorem}

By the equivalence of~\cite{KNR92}, we obtain the following.

\begin{corollary}\label{cor:universal}
For every fixed $k$ and every $n$ there is a graph with $n(\log^\ast n)^{O_k(1)}$ vertices that contains every $n$-vertex graph of treewidth at most $k$ as an induced subgraph.
\end{corollary}

For fixed $k\ge 2$, Theorem~\ref{thm:main} improves the second-order term of~\cite{GL07} from $O(\log\log n)$ to $O(\log\log^\ast n)$, and Corollary~\ref{cor:universal} improves the size of the induced-universal graph from $n(\log n)^{O(k)}$ to $n(\log^\ast n)^{O_k(1)}$. We stress that the dependence on $k$ is poor: the bound of Theorem~\ref{thm:main} is smaller than that of~\cite{GL07} only for $k=O((\log\log\log n)^{1/3})$.

\paragraph{What is new.} The previous labeling schemes for bounded treewidth~\cite{GL07,BGP20,DEGJMM21} identify each earlier neighbor of a vertex by its depth along a root path in a tree of logarithmic depth, at a cost of $O(\log\log n)$ bits per neighbor. We restate this in terms of laminar families of intervals: a vertex finds its earlier neighbor of color $c$ as the owner of the interval of color $c$ of a given nesting depth around it (Section~\ref{sec:prelim}). In this setting the logarithmic depth is forced by contiguity: if every subtree occupies consecutive positions and every vertex owns at most $R$ intervals, some trees of treewidth $3$ require depth $\Omega(\log n/\log R)$ (Proposition~\ref{prop:contig}). Our main conceptual contribution is to give up contiguity. We move every subtree that is heavy compared with the interval of its color around it, which reduces the depth to $4\log^\ast n+O_k(1)$, and we show that this is affordable. The correctness lost by the moves is restored with $2^{O(k^3)}$ intervals per vertex, and the bits needed to locate the moved intervals are paid for by the moved subtrees themselves: the subtrees moved out of one interval are disjoint and large, and along every root-to-leaf path they shrink fast enough.

\subsection{Our Methods}

Our labeling scheme combines two known techniques with two new ingredients. First, following the presentation in~\cite{BGP20} of the scheme of~\cite{GL07}, a vertex can identify each of its earlier neighbors by a depth along a root path; there, the path lies in a tree of depth at most $\log n$, so each depth takes $O(\log\log n)$ bits. We use nested intervals instead: a $k$-tree is represented by a tree in which every vertex is adjacent to its nearest ancestor of each other color, and every vertex stores, for each color, how many intervals of that color contain it (Section~\ref{sec:prelim}). The natural choice of intervals, the subtrees in depth-first order, works, but these numbers can be linear in $n$. Our first new ingredient reduces them to $4\log^\ast n+O_k(1)$: we move every heavy block of siblings out of the interval that contains it, and repair the intervals that a move breaks with additional intervals, at most $2^{O(k^3)}$ per vertex (Sections~\ref{sec:layout} and~\ref{sec:bounds}). Second, every vertex must compute where its intervals are. An additional interval lies near an interval of the same vertex that the vertex has already located, and can be found from it with $O(\log\log y)$ bits, where $y$ is the size of the surrounding region. To store these bits we use the approximate lengths and the identifiers aligned to powers of two of Alstrup, Dahlgaard and Knudsen~\cite{ADK17}, and hash functions with a few choices per key, in the spirit of $d$-ary cuckoo hashing~\cite{FPSS05}. Our second new ingredient is the accounting: we show that the identifiers wasted by the alignment are paid for by the subtrees of the moved blocks (Section~\ref{sec:labels}). Proofs that are not needed to follow the argument are deferred to the appendix.

\section{Preliminaries}\label{sec:prelim}

Logarithms are binary, and $\log^\ast n$ is the least $j\ge 0$ with $\log^{(j)}n\le 1$. Trees are rooted. For a vertex $v$ of a tree $T$ we write $p(v)$ for its parent, $T_v$ for the subtree rooted at $v$ and $s(v)=|T_v|$. Ancestors and descendants are proper unless stated otherwise. A coloring $\col:V(T)\to\{0,\dots,k\}$ is \emph{proper} if $\col(v)\ne\col(p(v))$ for every non-root $v$. For a vertex $x$ and a color $c$, $g_c(x)$ is the nearest proper ancestor of $x$ of color $c$, and $g_c(x)=\bot$ if there is none.

\subsection{Colored Trees}

\begin{lemma}\label{lem:colored}
Let $G$ be a $k$-tree on $n\ge k+1$ vertices. There are a rooted tree $T$ on $V(G)$ and a proper coloring $\col:V(G)\to\{0,\dots,k\}$ of $T$ such that
\[
E(G)=\bigl\{\{x,g_c(x)\}:x\in V(G),\ c\ne\col(x),\ g_c(x)\ne\bot\bigr\}.
\]
\end{lemma}
This representation follows from the elimination ordering of $k$-trees and their unique proper $(k+1)$-coloring (see, e.g.,~\cite{DEGJMM21}); for completeness, a proof is given in Appendix~\ref{app:prelim}.

Every graph of treewidth at most $k$ on at least $k+1$ vertices is a spanning subgraph of a $k$-tree~\cite{Bod98}. Given $T$ and $\col$ as in Lemma~\ref{lem:colored}, a vertex $x$ records in $k$ bits which of its edges to the vertices $g_c(x)$ belong to the subgraph. Hence it suffices to give labels from which the decoder decides, for two distinct vertices $a,b$, whether $a=g_{\col(a)}(b)$: applying this test in both directions and reading the recorded bit gives adjacency. Graphs on at most $k$ vertices are labeled with $O(k)$ bits directly.

Hereafter, we assume $k\ge 1$ to be fixed and $T$ to be a rooted tree on $n$ vertices with a proper coloring in $k+1$ colors. The vertices $g_c(x)\ne\bot$, $c\ne\col(x)$, are the \emph{upper neighbors} of $x$. One of them is $p(x)=g_{\col(p(x))}(x)$.

\subsection{Interval Families and Correctness}

\begin{definition}\label{def:rep}
An \emph{interval family} for $T$ consists of an injective map $\id:V(T)\to\mathbb Z$ and a laminar family $\mathcal F$ of non-empty integer intervals, each \emph{owned} by a vertex $o(I)$ and of \emph{color} $\col(I)=\col(o(I))$. It is \emph{correct} if for every vertex $x$ and every color $c\ne\col(x)$ with $g_c(x)\ne\bot$, the innermost member of $\mathcal F$ of color $c$ that contains $\id(x)$ is owned by $g_c(x)$. We write $\alpha_c(x)$ for the number of members of color $c$ that contain $\id(x)$, and the \emph{depth} of $I\in\mathcal F$ is the number of members of color $\col(I)$ that contain $I$, including $I$.
\end{definition}

\begin{lemma}[counting test]\label{lem:gatetest}
In a correct family, let $x$ be a vertex, $c\ne\col(x)$ a color with $g_c(x)\ne\bot$, and $o$ a vertex of color $c$. Then $o=g_c(x)$ if and only if $\id(x)$ lies in an interval of $o$ of depth $\alpha_c(x)$.
\end{lemma}
\begin{proof}
By laminarity the members of color $c$ containing $\id(x)$ form a chain $I_1\supsetneq\dots\supsetneq I_\alpha$ with $\alpha=\alpha_c(x)$, and a member of color $c$ containing $I_j$ contains $\id(x)$, so $I_j$ has depth $j$. Hence the only interval of color $c$ and depth $\alpha$ containing $\id(x)$ is $I_\alpha$, which is owned by $g_c(x)$, as desired.
\end{proof}

So a vertex $x$ stores $\id(x)$ and the numbers $\alpha_c(x)$, and an owner must be able to compute its intervals and their depths from its own label. If no interval of color $c$ contains $\id(x)$ and $g_c(x)=\bot$, the vertex stores $\alpha_c(x)=\bot$ and the test answers no.

Identifying an ancestor by a depth is not new. In the presentation in~\cite{BGP20} of the scheme of~\cite{GL07}, every vertex stores, for each of its at most $k$ earlier neighbors, the depth of the node containing it in a decomposition tree of depth at most $\log n$, which takes $O(\log\log n)$ bits per neighbor. The intervals of Section~\ref{sec:layout} have depth $O(\log^\ast n)$ instead.

\paragraph{Tools from~\cite{ADK17}.} We use two techniques of Alstrup, Dahlgaard and Knudsen. First, a $(1+\varepsilon)$-approximation of an integer $a$ can be stored in $O(\log\log a+\log(1/\varepsilon))$ bits. We round the length $y$ of every container of our layout, a set of consecutive positions, in this way with $\varepsilon=1/(\Phi\lambda(y)^4)$, where $\lambda(y)=\lfloor\log y\rfloor+2$ and $\Phi$ is a parameter, and we align the container to a power of two of about the rounding precision. Then a \emph{code} of $O(\log\lambda(y)+\log\Phi)$ bits describes the length, and the first position of the container can be recovered from any position inside it, the code, and $O(\log\lambda(y)+\log\Phi)$ further bits (Appendix~\ref{app:prelim}). Second, identifiers are aligned to powers of two. If $\id(x)\in[0,N)$ is a multiple of $2^z$, the label stores the Elias code $\gamma(z+1)$~\cite{Eli75}, then exactly $z-|\gamma(z+1)|$ bits of other data, and then $\id(x)$ without its $z$ trailing zeros, in $\lceil\log N\rceil-z$ bits. These parts take $\lceil\log N\rceil$ bits together and can be separated by the decoder; we say that the other data are \emph{absorbed}. Aligning $\id(x)$ to $2^z$ wastes fewer than $2^z$ identifiers, which Section~\ref{sec:labels} pays for.

\section{The Layout}\label{sec:layout}

\begin{figure}[t]
\centering
\begin{tikzpicture}[x=1cm,y=1cm]
\begin{scope}
\foreach \n/\c/\y in {r/cR/0,a/cB/-0.75,b/cG/-1.5,c/cB/-2.25,d/cR/-3.0,x/cG/-3.75}{
  \node[vtx=\c] (T\n) at (0,\y) {$\n$};
}
\draw[thick] (Tr)--(Ta)--(Tb)--(Tc)--(Td)--(Tx);
\draw[dashed,thick,gray] (Tb.east) .. controls +(0.55,0) and +(0.55,0) .. (Tr.east);
\draw[dashed,thick,gray] (Tc.east) .. controls +(1.15,0) and +(1.15,0) .. (Tr.east);
\draw[dashed,thick,gray] (Tx.east) .. controls +(0.55,0) and +(0.55,0) .. (Tc.east);
\draw[dashed,thick,gray] (Td.west) .. controls +(-0.55,0) and +(-0.55,0) .. (Tb.west);
\node[font=\small] at (0.25,-4.55) {(a)};
\end{scope}
\begin{scope}[shift={(2.9,0.3)}]
\foreach \n/\c/\xx in {r/cR/0,a/cB/0.8,b/cG/1.6,c/cB/2.4,d/cR/3.2,x/cG/4.0}{
  \node[vtx=\c] at (\xx,0) {$\n$};
}
\foreach \l/\a/\b/\c/\y in {{B(r)}/0.5/4.3/cR/-0.45,{B(a)}/1.3/4.3/cB/-0.8,{B(b)}/2.1/4.3/cG/-1.15,{B(c)}/2.9/4.3/cB/-1.5,{B(d)}/3.7/4.3/cR/-1.85}{
  \draw[bar=\c] (\a,\y)--(\b,\y) node[pos=0,left=3pt,black,font=\scriptsize] {$\l$};
}
\node[font=\small] at (-0.6,0) {(b)};
\end{scope}
\begin{scope}[shift={(2.9,-2.25)}]
\foreach \n/\c/\xx in {r/cR/0,a/cB/0.8,b/cG/1.6,c/cB/2.4,d/cR/3.2,x/cG/4.0}{
  \node[vtx=\c] at (\xx,0) {$\n$};
}
\foreach \l/\a/\b/\c/\y in {{B(r)}/0.5/4.3/cR/-0.45,{B(a)}/1.3/2.7/cB/-0.8,{B(b)}/2.1/2.7/cG/-1.15,{B(c)}/2.9/4.3/cB/-1.85,{B(d)}/3.7/4.3/cR/-2.2}{
  \draw[bar=\c] (\a,\y)--(\b,\y) node[pos=0,left=3pt,black,font=\scriptsize] {$\l$};
}
\draw[wbar=cG] (2.9,-1.5)--(4.3,-1.5) node[pos=0,left=3pt,black,font=\scriptsize] {wrap of $b$};
\draw[gray,rounded corners=2pt,densely dotted,thick] (2.84,-1.3) rectangle (4.42,-2.38);
\node[gray,font=\scriptsize,anchor=west] at (4.45,-1.85) {area of $B(a)$};
\node[font=\small] at (-0.6,0) {(c)};
\end{scope}
\begin{scope}[shift={(11.0,0.3)},x=0.85cm,y=0.78cm,
  iv/.style={draw=#1,line width=1.1pt,rounded corners=2pt,inner sep=2pt,font=\scriptsize,fill=#1!8},
  wv/.style={draw=#1,line width=1.1pt,dash pattern=on 2.5pt off 1.5pt,rounded corners=2pt,inner sep=2pt,font=\scriptsize,fill=#1!8},
  av/.style={draw=gray,densely dotted,thick,rounded corners=2pt,inner sep=2pt,font=\scriptsize}]
\node[font=\scriptsize] (R) at (0,0) {root};
\node[vtx=cR] (r) at (-1.3,-0.9) {$r$};
\node[iv=cR] (Br) at (1.0,-0.9) {$B(r)$};
\node[vtx=cB] (a) at (-0.7,-1.8) {$a$};
\node[iv=cB] (Ba) at (0.6,-1.8) {$B(a)$};
\node[vtx=cG] (b) at (-0.1,-2.7) {$b$};
\node[iv=cG] (Bb) at (1.2,-2.7) {$B(b)$};
\node[vtx=cB] (c) at (1.2,-3.6) {$c$};
\node[av] (A) at (3.0,-1.8) {area};
\node[wv=cG] (Wb) at (3.0,-2.7) {wrap of $b$};
\node[iv=cB] (Bc) at (3.0,-3.6) {$B(c)$};
\node[vtx=cR] (d) at (2.4,-4.5) {$d$};
\node[iv=cR] (Bd) at (3.6,-4.5) {$B(d)$};
\node[vtx=cG] (x) at (3.6,-5.4) {$x$};
\draw (R)--(r); \draw (R)--(Br); \draw (Br)--(a); \draw (Br)--(Ba); \draw (Br)--(A);
\draw (Ba)--(b); \draw (Ba)--(Bb); \draw (Bb)--(c);
\draw (A)--(Wb); \draw (Wb)--(Bc); \draw (Bc)--(d); \draw (Bc)--(Bd); \draw (Bd)--(x);
\node[font=\small] at (-1.3,0) {(d)};
\end{scope}
\end{tikzpicture}
\caption{A small example with $k=2$ and the colors red, blue and green. (a)~A colored tree. Every vertex is adjacent to its nearest ancestor of each other color; the dashed lines are the edges that are not tree edges. For instance, $x$ is adjacent to $d$ and $c$. (b)~The blocks in depth-first order. The blue intervals around $x$ are $B(a)$ and $B(c)$, and the inner one is owned by $c$, the blue neighbor of $x$. (c)~The block $B(c)$ is moved out of its home $B(a)$ into the area of $B(a)$, right after it. Now $x$ lies in a single blue interval. The move took $d$ out of the green block $B(b)$, although $b$ is the green neighbor of $d$, so $B(c)$ is surrounded by a wrap of $b$, whose source is $B(b)$. (d)~The layout tree of (c); its leaves, read from left to right, give the order of (c).}\label{fig:running}
\end{figure}
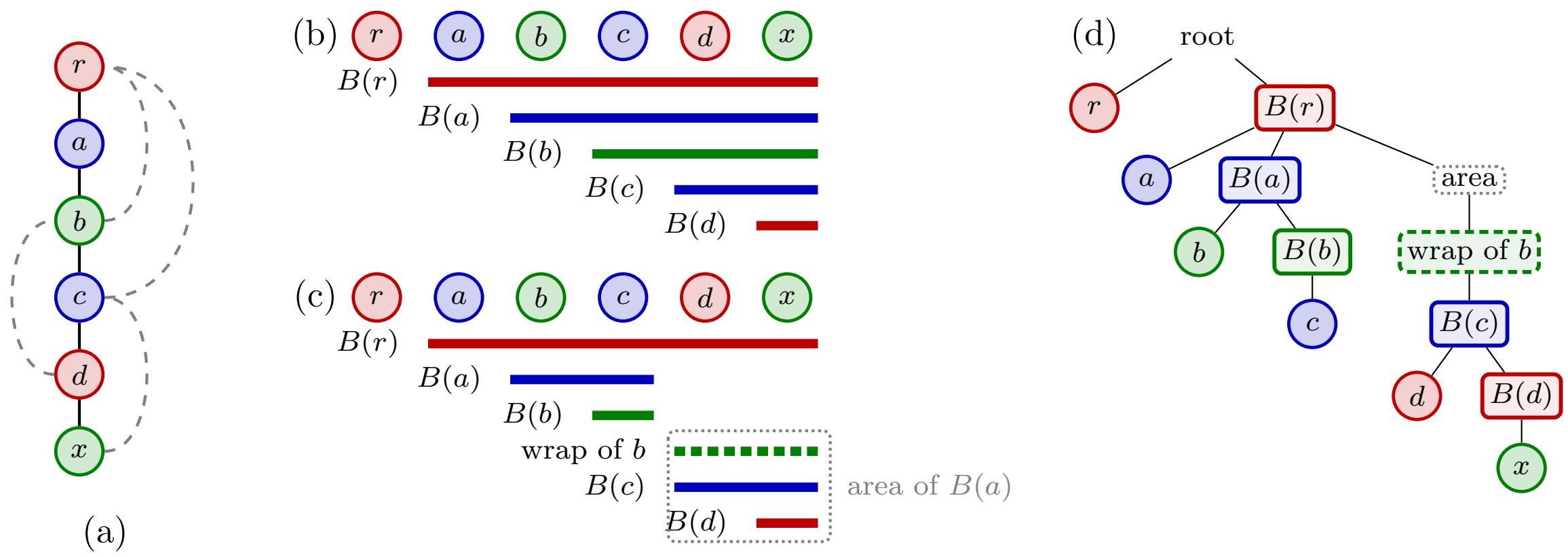

In this section we construct the order and the intervals. We start with blocks, which give a correct family whose depth can be large. We then move heavy blocks to reduce the depth, and repair correctness with additional intervals, called wraps. Figure~\ref{fig:running} follows the construction on a small example.

\subsection{Blocks}

\begin{definition}\label{def:block}
For a vertex $v$ with at least one child, the \emph{block} $B(v)$ is the set of children of $v$. As an interval it has color $\col(v)$ and owner $v$, and its \emph{mass} is $\mass(B(v))=s(v)-1$.
\end{definition}

\begin{lemma}\label{lem:blockpath}
Let $x$ be a vertex, $c$ a color with $g=g_c(x)\ne\bot$, and $x=y_0,y_1,\dots,y_M=g$ the tree path. Then no $y_i$ with $1\le i<M$ has color $c$, and $x$ lies in the subtree of $y_{M-1}\in B(g)$.
\end{lemma}
\begin{proof}
This is the definition of $g_c$.
\end{proof}

If the vertices are written in depth-first order and every block is written as the interval of the subtrees of its members, then by Lemma~\ref{lem:blockpath} the innermost interval of color $c$ around $x$ is $B(g_c(x))$, so this family is correct (Figure~\ref{fig:running}(b)). Its depth, however, can be linear: if $T$ is a path colored $0,1,\dots,k,0,1,\dots$, the blocks of one color are nested about $n/(k+1)$ deep. The following proposition, proved in Appendix~\ref{app:contig}, shows that no layout in which subtrees occupy consecutive positions does much better, even if a vertex owns several intervals.

\begin{proposition}\label{prop:contig}
For all $R\ge 1$ and $d\ge 0$ there is a colored tree $U_d$ with $k=3$ and at most $11(2R+1)^d$ vertices such that every correct family for $U_d$ in which every subtree occupies consecutive positions and every vertex owns at most $R$ intervals has depth at least $d+1$. Hence the depth is at least $\log n/\log(2R+1)-O(1)$.
\end{proposition}

\subsection{Moving Blocks and Repairing Correctness}

We keep a block in its place only if it is light. The \emph{home} of $B(v)$ is the innermost interval $J$ of color $\col(v)$ around $v$. The block is \emph{light} if its mass is smaller than $\tau(\sigma(J))$, where $\tau(s)\approx\lceil\log s\rceil^\beta$ is a threshold and $\sigma(J)$ is, roughly, the size of the subtree of the owner of $J$. A heavy block is \emph{moved}: it is written right after its home $J$, outside $J$. If $J$ is kept, the moved block goes to the \emph{area} of $J$, an unowned stretch of positions right after $J$; if $J$ itself lies in an area, it goes next to $J$, in the same area (Figure~\ref{fig:running}(c)). As a result, two nested blocks of one color are always light relative to each other, so their masses drop from $s$ to below $\tau(s)$ every two levels of nesting, and the depth is $O(\log^\ast n)$.

Moving a block, however, breaks correctness for the other colors. Before the move, the vertices in $B(v)$ were inside the intervals of all ancestors of $v$, and afterwards they are inside only those that also contain the new place of $B(v)$. In Figure~\ref{fig:running}(c), the vertex $d$ is no longer inside the green block of $b$, although $b$ is its green neighbor. We therefore surround the moved block with a \emph{wrap} of $b$, an interval owned by $b$. In general, $B(v)$ gets a wrap of $g_c(v)$ for every color $c$ for which its new place is not already inside an interval of $g_c(v)$ of the right nesting level. The wrap replaces the interval of $g_c(v)$ around $v$ before the move, which we call its \emph{source}. Moved blocks that need the same wraps share them, so the number of wraps stays small.

\subsection{The Layout Tree}

We describe the layout as a rooted tree $\cL$, the \emph{layout tree}, whose leaves are the vertices of $T$ and whose inner nodes are of four kinds: the root, blocks, wraps and areas (Figure~\ref{fig:running}(d)). Blocks and wraps are \emph{intervals}; each has an owner and the color of its owner. Any depth-first order of $\cL$ lists the leaves below a node consecutively, so with $\id(x)$ the position of $x$ in such an order, every interval becomes an integer interval, and the intervals form a laminar family. Section~\ref{sec:labels} chooses the order and inserts gaps. For a node $N$, its \emph{interval ancestors} are the intervals on the path from $N$ to the root, $N$ included if it is an interval. ``Below'' means ``a descendant of'' in $\cL$, and $o(L)$ is the owner of an interval $L$.

Fix a nondecreasing function $\tau\ge 1$, the \emph{threshold}. Every block is either \emph{kept} or \emph{moved}, and every moved block and every wrap belongs to an area. For an interval $L$ let $\sigma(L)=s(o(L))$ if $L$ is a kept block, and $\sigma(L)=s(o(J_A))$ if $L$ belongs to the area $A$, where $J_A$ is the kept block whose area is $A$. The layout tree is built by $\emitf(\mathrm{root},\mathrm{root\ of\ }\cL)$, where $\emitf(v,N)$ does the following.
\begin{enumerate}[label=(E\arabic*)]
\item Attach a leaf $v$ as a child of $N$. If $v$ has no children, stop.
\item Let $J$ be the nearest interval ancestor of the leaf $v$ of color $\col(v)$, or $\bot$. If $J=\bot$ or $\mass(B(v))<\tau(\sigma(J))$, attach $B(v)$ as a kept block, as a child of $N$, call $\emitf(w,B(v))$ for every child $w$ of $v$, and stop.
\item Otherwise $B(v)$ is moved, and $J$ is its \emph{home}. If $J$ is a kept block, let $A$ be the \emph{area of $J$}, a child of the parent of $J$ created when first needed, and let $P=A$. Otherwise $J$ belongs to an area $A$, and let $P$ be the parent of $J$.
\item For every color $c\ne\col(v)$ with $g_c(v)\ne\bot$, the moved block \emph{needs} $z=g_c(v)$ if the nearest interval ancestor of $P$ of color $c$ is not owned by $z$. Let $z_1,\dots,z_m$ be the needed vertices, ordered from the root down, and let the \emph{source} $S_i$ be the nearest interval ancestor of the leaf $v$ of color $\col(z_i)$.
\item Starting at $N_0=P$, for $i=1,\dots,m$ let $N_i$ be the child of $N_{i-1}$ that is a wrap with owner $z_i$ and source $S_i$, created if it does not exist; it belongs to $A$ and has home $J$. Attach $B(v)$ as a child of $N_m$, and call $\emitf(w,B(v))$ for every child $w$ of $v$.
\end{enumerate}
A wrap is created for the first moved block that passes through it, and later moved blocks with the same $P$ and the same pairs $(z_i,S_i)$ reuse it. Hence every wrap has one owner, one source and one home. The procedure is a depth-first traversal of $T$ that never changes the parent of a node, so the interval ancestors of a leaf are fixed when it is attached, and the layout tree is well defined.

\section{Correctness and Depth}\label{sec:bounds}

\subsection{Correctness}

The following theorem is the heart of the construction. Its proof follows the tree path from a vertex $x$ up to $g=g_c(x)$. The blocks on the way have colors other than $c$, and as long as they are kept, the path in the layout tree simply climbs through them. At the first moved block on the way, either its new place already lies in an interval of color $c$ owned by $g$, or a wrap of $g$ surrounds it.

\begin{theorem}\label{thm:exact}
For every vertex $x$ and every color $c$ with $g=g_c(x)\ne\bot$, the nearest interval ancestor of the leaf $x$ of color $c$ is owned by $g$. In particular the family of intervals is correct, whatever the threshold.
\end{theorem}
\begin{proof}
Let $x=y_0,\dots,y_M=g$ be the tree path; by Lemma~\ref{lem:blockpath}, $\col(y_i)\ne c$ for $1\le i<M$. The leaf $y_{i-1}$ is a child of $B(y_i)$. If $B(y_i)$ is kept, it is a sibling of the leaf $y_i$, hence a child of $B(y_{i+1})$. Let $i^\ast$ be the least $i\in[1,M-1]$ such that $B(y_i)$ is moved, if there is one. Then the path from the leaf $x$ up to $B(y_{i^\ast})$, or up to $B(g)$ if there is no such $i$, passes only through the blocks $B(y_1),\dots,B(y_{i^\ast-1})$, whose colors differ from $c$. In the second case we are done. In the first, $\col(y_{i^\ast})\ne c$ and $g_c(y_{i^\ast})=g$. Let $P$ and $N_1,\dots,N_m$ be as in (E3)--(E5) for $B(y_{i^\ast})$; the wraps $N_i$ have distinct colors. If the nearest interval ancestor of $P$ of color $c$ is owned by $g$, then $g$ is not needed and no $N_i$ has color $c$. Otherwise $g$ is needed, and its wrap is the only $N_i$ of color $c$. In both cases the nearest interval of color $c$ above $B(y_{i^\ast})$ is owned by $g$, as desired.
\end{proof}

Note that for colors $c\ne\col(x)$ this is correctness, while for $c=\col(x)$ it says that the home of a moved block $B(x)$ is owned by $g_{\col(x)}(x)$. In Appendix~\ref{app:layout} we derive further properties of the layout. The most important one is \emph{lightness}: every block $R$ of color $c$ has $\mass(R)<\tau(\sigma(K))$, where $K$ is the nearest interval of color $c$ strictly above $R$ (Lemma~\ref{lem:sigma}(c)). For kept blocks this is the rule (E2). For moved blocks it follows because the block of a needed vertex lies inside every interval of its color around its wraps (Lemma~\ref{lem:needed}).

\subsection{Depth}

For a nondecreasing $\tau$ let $m_\tau=\max\{m\ge 1:\tau(m)\ge m\}$, and let $L_\tau(n)$ be the least $L$ with $\tau^{L}(n)\le m_\tau$.

\begin{theorem}\label{thm:depth}
Every leaf lies in at most $2L_\tau(n)+m_\tau+1$ blocks and at most $2L_\tau(n)+m_\tau+1$ wraps of each color.
\end{theorem}

\noindent\emph{Proof idea.} If a block $R'$ lies strictly below a block $R$, then $\sigma(R')\le\mass(R)$, since the kept block that $R'$ belongs to lies below $R$. Let $R_1\supsetneq R_2\supsetneq\cdots$ be the blocks of one color around a point, with masses $x_1>x_2>\cdots$. By lightness, $x_{i+1}<\tau(\sigma(R_i))\le\tau(x_{i-1})$, so the masses drop by $\tau$ every two levels. The masses of the blocks of the owners of nested wraps of one color satisfy the same recursion. The full proof is given in Appendix~\ref{app:depth}.\hfill$\square$

\medskip
We use the threshold $\tau_\beta(s)=\max\{1,\min\{\lceil\log s\rceil^\beta,\lceil s/2\rceil\}\}$, where $\beta=2^{O(k^3)}$ is fixed in Appendix~\ref{app:waste}. It is nondecreasing, and $\tau_\beta(s)<s$ for $s\ge 2$, so $m_{\tau_\beta}=1$.

\begin{lemma}\label{lem:iter}
$L_{\tau_\beta}(n)\le\log^\ast n+3\beta\log(8\beta)$.
\end{lemma}

\begin{corollary}\label{cor:depth}
With $\tau=\tau_\beta$, every point lies in at most $D_n=4\log^\ast n+12\beta\log(8\beta)+4$ intervals of each color, and an interval depth is stored in $\log\log^\ast n+O(\log\beta)$ bits.
\end{corollary}

\subsection{Intervals per Vertex}

A vertex $z$ owns its block and its wraps. Every wrap of $z$ has a source, which is another interval of $z$, so the intervals of $z$ form a tree under the relation ``is the source of'', rooted at $B(z)$. The owner computes its intervals along this tree, from $B(z)$ down.

\begin{theorem}\label{thm:records}
Every vertex owns at most $\rho_k=2((k+1)2^k)^{(k+1)^2}=2^{O(k^3)}$ intervals.
\end{theorem}

\begin{lemma}\label{lem:depths}
Let $W$ be a wrap of $z$ with source $S$. Then the depth of $W$ is the depth of $S$ minus $\delta\in\{0,1\}$.
\end{lemma}

\noindent\emph{Proof idea.} A wrap $W$ of $z$ with source $S$ is determined by $S$, the color of the moved block for which it was created, and the set of colors of the vertices needed before $z$, so $S$ is the source of at most $k2^k$ wraps. Along a chain of sources the area changes at most $k$ times, since the homes at which it changes are nested and have distinct colors, and within one area the number of wraps above the current one strictly decreases. Hence chains of sources have length at most $(k+1)^2$. For the depths, lightness implies that at most one interval of color $\col(z)$ lies strictly between the home of $W$ and its source. The proofs are given in Appendix~\ref{app:records}.\hfill$\square$

\section{Labels}\label{sec:labels}

From now on $\tau=\tau_\beta$, $D_n$ is the bound of Corollary~\ref{cor:depth}, $w=\lceil\log(D_n+2)\rceil$, and $\Phi=8(k+2)2^{w}=O(k\log^\ast n)$. Every point lies in at most $(k+1)D_n\le\Phi/8$ intervals.

\subsection{Locating Intervals}

\begin{figure}[t]
\centering
\begin{tikzpicture}[x=1cm,y=1cm]
\draw[thick] (0,0) rectangle (7.2,0.55);
\node[font=\scriptsize,anchor=north west] at (0.0,-0.02) {home $J$};
\draw[thick,densely dashed,gray] (2.0,0.08) rectangle (5.6,0.47);
\draw[line width=2.4pt,cG] (3.0,0.27)--(4.4,0.27);
\node[font=\scriptsize,cG,anchor=north] at (3.7,-0.02) {source $S$ of $z$};
\draw[thick] (7.4,0) rectangle (13.6,0.55);
\node[font=\scriptsize,anchor=south] at (10.5,0.6) {area of $J$: sub-areas, each with slots of one length};
\foreach \a/\b/\s in {7.55/8.85/0.26,9.05/11.05/0.4,11.25/13.45/0.55}{
  \draw[gray] (\a,0.08) rectangle (\b,0.47);
  \pgfmathsetmacro{\st}{\a+\s}
  \pgfmathsetmacro{\stt}{\a+2*\s}
  \foreach \t in {\st,\stt,...,\b}{ \draw[gray!70] (\t,0.08)--(\t,0.47); }
}
\fill[cG!25] (12.35,0.09) rectangle (12.9,0.46);
\draw[wbar=cG] (12.4,0.27)--(12.85,0.27);
\node[font=\scriptsize,cG,anchor=west] at (13.65,0.27) {wrap $W$ of $z$};
\tikzset{num/.style={circle,draw,fill=white,inner sep=0.6pt,font=\scriptsize}}
\draw[->,thick] (3.0,0.5) to[out=140,in=40] node[pos=0.5,num] {1} (0.05,0.5);
\draw[->,thick] (7.2,0.3) -- node[pos=0.5,num,yshift=7pt] {2} (7.45,0.3);
\draw[->,thick] (7.5,-0.12) to[out=-30,in=-150] node[pos=0.5,num] {3} (11.25,-0.12);
\draw[->,thick] (11.3,-0.45) to[out=-10,in=-100] node[pos=0.5,num] {4} (12.62,0.05);
\end{tikzpicture}
\caption{How the owner $z$ of a wrap $W$ finds it. The source $S$ of $W$ is an interval of $z$ that $z$ has already located, and it lies inside the home $J$. (1) The owner recovers the start of $J$ from the position of $S$ and a few bits, since $J$ is aligned. (2) The area starts right after $J$. (3) The owner jumps to the sub-area whose slots have the length of $W$. (4) It picks the slot of $W$ with a slot graph, using the position of $S$ as the key and a short choice stored in its label. When the walk has to pass through several wraps, each step is of the same kind. All of this happens inside the region of $J$ and costs $O(k\log\log y+k\log\log^\ast n)$ bits, where $y$ is the size of the region.}\label{fig:locate}
\end{figure}
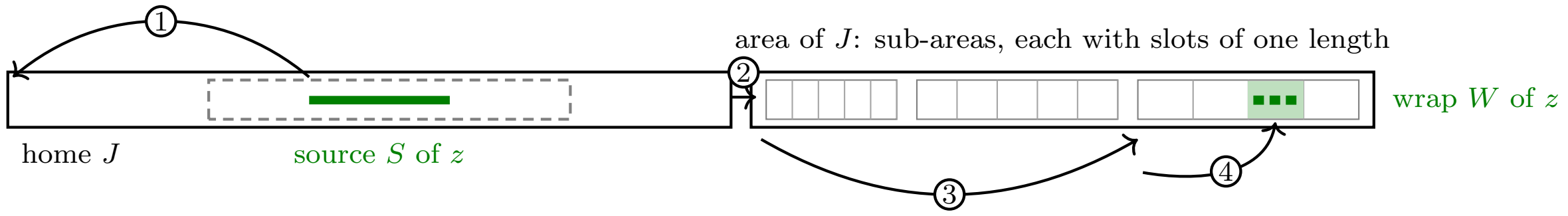

A vertex knows where its block starts if the block is kept, since it starts right after the vertex. It must, however, compute the positions of a moved block and of its wraps. The key observation is that a wrap is never far from its source: the source lies inside the home $J$, and the wrap lies right after $J$ (Figure~\ref{fig:locate}). The owner therefore walks from its source up to $J$ and over to the wrap. Each step up recovers the start of an aligned container from a position inside it, which takes $O(\log\lambda(y)+\log\Phi)$ bits for a container of length $y$. Each step down picks a slot by means of a \emph{slot graph}, a hash function with a few choices per key in the spirit of $d$-ary cuckoo hashing~\cite{FPSS05}. Appendix~\ref{app:bits} describes the placement in detail.

Every vertex $v$ gets an \emph{alignment} $z(v)\ge 0$, chosen below, and $\id(v)$ is a multiple of $2^{z(v)}$. For a home $J$, the \emph{region} $\mathrm{Reg}(J)$ consists of $J$ and the containers placed right after $J$ that hold the moved blocks and wraps with home $J$; $y(J)$ is its length. The set of vertices in $\mathrm{Reg}(J)$ does not depend on the alignments, and we let $\hat y(J)=2^{10}\sum_u 2^{z(u)}$, the sum over the vertices $u$ in $\mathrm{Reg}(J)$. The \emph{key} of a moved block is the leaf of its owner, and the key of a wrap is its source. The following two lemmas are proved in Appendix~\ref{app:bits}.

\begin{lemma}[locating]\label{lem:locate}
Let $U$ be a moved block or a wrap with home $J$. There are data $\mathrm{loc}(U)$ of $O(k(\log\lambda(y(J))+\log\Phi+\log k))$ bits such that the first position of $U$ is determined by $\mathrm{loc}(U)$ and the first and last positions of the key of $U$.
\end{lemma}

\begin{lemma}[size]\label{lem:size}
Let $W=\sum_v 2^{z(v)}$. Then $N\le e^6(n+W)$, and $y(J)\le\hat y(J)$ for every home $J$.
\end{lemma}

The alignments determine the layout, and the layout determines the alignments; Appendix~\ref{app:bits} shows that the rule below has a fixed point, which we use.

\subsection{The Encoder and the Decoder}

Let $R$ be an interval of $v$. Its home is the home of $R$ if $R$ is a moved block or a wrap, and $R$ itself if $R=B(v)$ is kept. The \emph{data} of $R$ are: for $B(v)$, whether it is kept and the code of its length; for a moved block or a wrap, the code of its length and $\mathrm{loc}(R)$; for a wrap, also the index of its source among the intervals of $v$ and the offset $\delta$ of Lemma~\ref{lem:depths}. They are written in a self-delimiting way and, by Lemmas~\ref{lem:locate} and~\ref{lem:size} and Theorem~\ref{thm:records}, take at most
\[
F_R=C_1\bigl(k\lceil\log\lambda(\hat y(J))\rceil+k\lceil\log\Phi\rceil+k^3\bigr)
\]
bits, for an absolute constant $C_1$, where $J$ is the home of $R$. We call $R$ \emph{small} if $\lambda(\hat y(J))\le 4Z$, where $Z=2^{O(k^3)}(1+\log\Phi)$ is fixed in Appendix~\ref{app:waste}. The data of the small intervals are stored explicitly, and the data of the other intervals of $v$ are absorbed in the identifier. Let $F_v$ be the sum of $F_R$ over the intervals that are not small. The alignment $z(v)$ is the least $z\ge 0$ with $z\ge|\gamma(z+1)|+F_v$.

The label $\ell(v)$ of a vertex $v$ consists of:
\begin{itemize}
\item a head: $\gamma(w+1)$; $\col(v)$; for every color $c\ne\col(v)$ the value $\alpha_c(v)$ in $w$ bits, or a code for $\bot$ if $g_c(v)=\bot$; if $v$ has children, the depth of $B(v)$ in $w$ bits; and the $k$ bits recording which edges to the vertices $g_c(v)$ are present;
\item the number of intervals of $v$, and for each of them, in the order of the tree of sources, whether it is small;
\item the data of the small intervals;
\item the identifier in absorbed form: $\gamma(z(v)+1)$, the data of the other intervals padded to $z(v)-|\gamma(z(v)+1)|$ bits, and $\lfloor\id(v)/2^{z(v)}\rfloor$ in $\lceil\log N\rceil-z(v)$ bits.
\end{itemize}

Given $\ell(a)$ and $\ell(b)$, the decoder computes $\Phi$ from $w$ and parses both labels. It decides whether $a=g_{\col(a)}(b)$ as follows. If $\col(a)=\col(b)$ or $\alpha_{\col(a)}(b)=\bot$, the answer is no. Otherwise it computes the intervals of $a$ along the tree of sources: the position of $B(a)$, which starts at the least multiple of its grain after $\id(a)$ if it is kept, and is found by Lemma~\ref{lem:locate} from $\id(a)$ otherwise; and the position of every wrap from its source by Lemma~\ref{lem:locate}. The lengths come from the codes, the depth of $B(a)$ from the head, and the depth of a wrap is that of its source minus $\delta$ by Lemma~\ref{lem:depths}. The answer is yes if and only if $\id(b)$ lies in an interval of $a$ of depth $\alpha_{\col(a)}(b)$. Recall from Section~\ref{sec:prelim} that applying this test in both directions and reading the edge bit of the lower vertex decides adjacency.

\begin{theorem}\label{thm:decoder}
The decoder is correct.
\end{theorem}
\begin{proof}
The decoder computes the positions and depths of all intervals of $a$ correctly by Lemmas~\ref{lem:locate} and~\ref{lem:depths}. The family is correct by Theorem~\ref{thm:exact}, so Lemma~\ref{lem:gatetest} applies.
\end{proof}

\subsection{The Cost of Alignment}

It remains to bound $W=\sum_v 2^{z(v)}$, the number of identifiers used once every vertex is aligned.

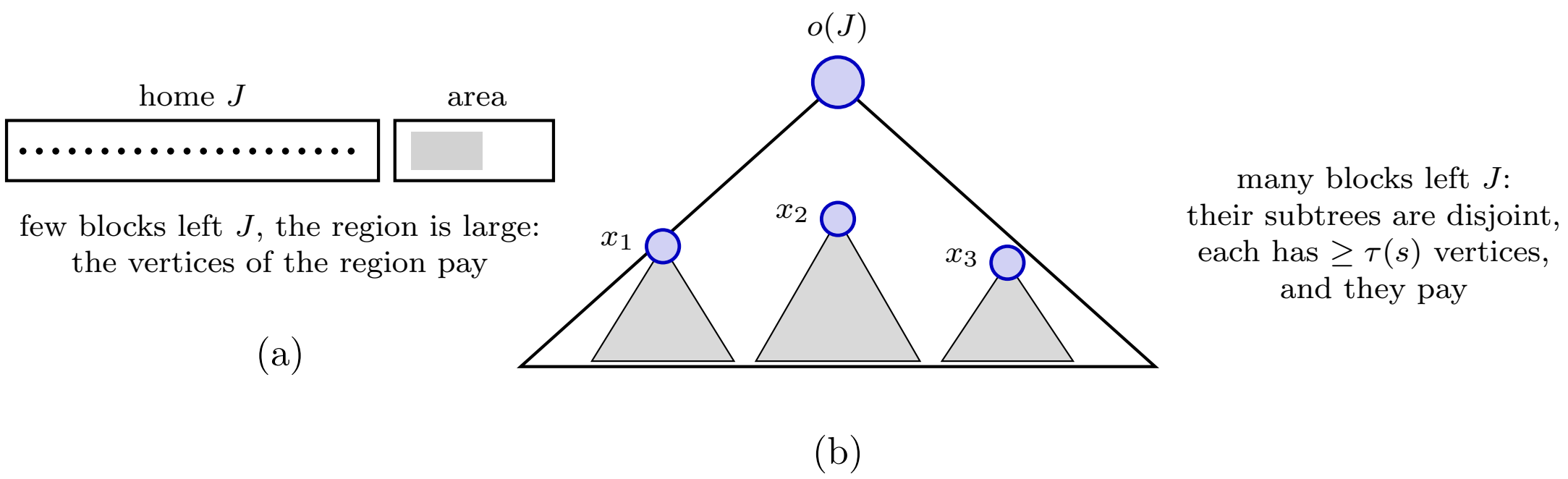
\begin{figure}[t]
\centering
\begin{tikzpicture}[x=1cm,y=1cm]
\begin{scope}
\draw[thick] (0,0) rectangle (3.4,0.55);
\node[font=\scriptsize,anchor=south] at (1.7,0.55) {home $J$};
\foreach \t in {0.15,0.3,...,3.3}{ \fill (\t,0.27) circle (0.9pt); }
\draw[thick] (3.55,0) rectangle (5.0,0.55);
\node[font=\scriptsize,anchor=south] at (4.3,0.55) {area};
\fill[gray!35] (3.7,0.1) rectangle (4.35,0.45);
\node[font=\scriptsize,align=center] at (2.5,-0.6) {few blocks left $J$, the region is large:\\ the vertices of the region pay};
\node at (2.5,-1.6) {\small (a)};
\end{scope}
\begin{scope}[shift={(7.6,0.9)}]
\draw[thick] (0,0)--(-2.9,-2.6)--(2.9,-2.6)--cycle;
\node[vtx=cB] (oJ) at (0,0) {};
\node[font=\scriptsize,anchor=south] at (0,0.22) {$o(J)$};
\foreach \px/\py/\w/\h/\lab in {-1.6/-1.5/0.65/1.05/{x_1},0.0/-1.25/0.75/1.3/{x_2},1.55/-1.65/0.6/0.9/{x_3}}{
  \fill[gray!30] (\px,\py)--({\px-\w},{\py-\h})--({\px+\w},{\py-\h})--cycle;
  \draw (\px,\py)--({\px-\w},{\py-\h})--({\px+\w},{\py-\h})--cycle;
  \node[vtx=cB,minimum size=0.3cm] at (\px,\py) {};
  \node[font=\scriptsize,anchor=east] at ({\px-0.12},{\py+0.05}) {$\lab$};
}
\node[font=\scriptsize,align=center] at (4.9,-1.4) {many blocks left $J$:\\ their subtrees are disjoint,\\ each has $\ge\tau(s)$ vertices,\\ and they pay};
\node at (0,-2.5-0.9) {\small (b)};
\end{scope}
\end{tikzpicture}
\caption{Who pays for the bits hidden in the identifiers of the owners of the moved blocks and wraps with home $J$. (a) If the region of $J$ has many vertices compared with the number of blocks moved out of $J$, the vertices of the region pay. (b) Otherwise at least two blocks $B(x_1),B(x_2),\dots$ left $J$. Their owners are not below one another, since the home of each of them is owned by $g_{\col(x_i)}(x_i)$, so their subtrees are disjoint, and each has at least $\tau(s)\ge(\log s)^\beta$ vertices, where $s$ is the subtree size of $o(J)$.}\label{fig:pay}
\end{figure}

\begin{lemma}\label{lem:waste}
$W\le 2^{2^{O(k^3)}}\Phi^{2^{O(k^3)}}\,n$.
\end{lemma}

\noindent\emph{Proof idea.} A moved block may retain few of its vertices, since blocks below it may have been moved further out, so its cost cannot be charged to the vertices it contains. Instead, every vertex that absorbs data is charged to the home $J$ of its largest absorbed interval (Figure~\ref{fig:pay}). If the region of $J$ has many vertices compared with the number of moved blocks with home $J$, the vertices of the region pay; every vertex lies in few regions. Otherwise at least two blocks were moved out of $J$. By Theorem~\ref{thm:exact}, the home of a moved block $B(x)$ is owned by $g_{\col(x)}(x)$, so no moved block with home $J$ has its owner below another one, and their subtrees are disjoint. Each has at least $\tau(\sigma(J))=\lceil\log\sigma(J)\rceil^\beta$ vertices, and these subtrees pay. Along a root-to-leaf path of $T$, the masses $m_j$ of consecutive such blocks of one color satisfy $m_j\ge m_{j+1}+\tau(\sigma_{j+1})$, which bounds the total charge per vertex by $\mathrm{poly}(\Phi)$ once $\beta$ is large compared with the number of bits per interval. The full proof is given in Appendix~\ref{app:waste}.\hfill$\square$

\begin{proof}[Proof of Theorem~\ref{thm:main}]
By Section~\ref{sec:prelim} it suffices to label colored trees, and the decoder is correct by Theorem~\ref{thm:decoder}. Since $\beta=2^{O(k^3)}$, we have $\log\Phi=w+O(\log k)=\log\log^\ast n+O(k^3)$ by Corollary~\ref{cor:depth}. The absorbed form of the identifier takes $\lceil\log N\rceil$ bits, and $N\le e^6(n+W)\le 2^{2^{O(k^3)}}\Phi^{2^{O(k^3)}}n$ by Lemmas~\ref{lem:size} and~\ref{lem:waste}, so $\lceil\log N\rceil\le\log n+2^{O(k^3)}(\log\log^\ast n+1)$. The head has $O(k(\log\log^\ast n+\log\beta))$ bits. A small interval has data of $C_1(k\log(4Z)+k\log\Phi+k^3)$ bits, and there are at most $\rho_k=2^{O(k^3)}$ intervals per vertex by Theorem~\ref{thm:records}, so the explicit part has $2^{O(k^3)}(\log\log^\ast n+1)$ bits. In total the labels have $\log n+2^{O(k^3)}(\log\log^\ast n+1)=\log n+O(\log\log^\ast n)$ bits for fixed $k$, as desired.
\end{proof}

\section{Concluding Remarks}\label{sec:open}

We have devised a labeling scheme of size $\log n+O(\log\log^\ast n)$ for graphs of bounded treewidth. Two parts of our labels grow with $n$: the $k+1$ depths of at most $4\log^\ast n+O_k(1)$, and the precision $1/\Phi$ of the alignments, where $\Phi=O(k\log^\ast n)$, which prevents rounding errors from accumulating along the intervals around a point. Each costs $O(\log\log^\ast n)$ bits. By Proposition~\ref{prop:contig}, depth $O(1)$ requires moving almost every subtree; the threshold $\tau\equiv 1$ does so, but then the moved blocks are too light to pay for locating them. The question of Gavoille and Labourel~\cite{GL07}, whether a labeling scheme of size $\log n+O(k)$ exists, remains open for every $k\ge 2$, and a positive answer would have to remove both sources of $\log\log^\ast n$. It would also be of interest to obtain a polynomial dependence on $k$, which needs fewer intervals per vertex or a way to locate the intervals of one vertex jointly, and to replace the slot graphs, which we show to exist by counting, by an explicit construction, which would make encoding and decoding efficient.

\section*{Acknowledgments}
The author used Claude, a large language model by Anthropic, for assistance in developing and checking parts of the arguments and in drafting the text and figures. The author has verified all proofs and takes full responsibility for the content of this paper.

\clearpage
\appendix

\section{Proofs and Details for Section~\ref{sec:prelim}}\label{app:prelim}

\begin{proof}[Proof of Lemma~\ref{lem:colored}]
Color $G$ properly with $k+1$ colors; for a $k$-tree this coloring exists and is unique up to renaming. Fix a construction order: an initial $(k+1)$-clique, and then vertices $x$ each joined to a $k$-clique $K(x)$ of earlier vertices. Make the initial clique a path in creation order, and make every later vertex $x$ a child of the last created vertex $p$ of $K(x)$. We show by induction that the earlier neighbors of every vertex $x$ are the vertices $g_c(x)\ne\bot$, $c\ne\col(x)$. For the initial clique this is immediate. For a later $x$, the vertices of $K(x)\setminus\{p\}$ are earlier neighbors of $p$, so by induction they are vertices $g_c(p)$. The colors of $K(x)$ are all colors except $\col(x)$, so $K(x)\setminus\{p\}=\{g_c(p):c\notin\{\col(x),\col(p)\}\}=\{g_c(x):c\notin\{\col(x),\col(p)\}\}$, and $p=g_{\col(p)}(x)$.
\end{proof}

\paragraph{Codes, grains and residues.} For an integer $y\ge 1$ let $\ell=\lfloor\log y\rfloor$ and $\lambda(y)=\ell+2$, and let $\gamma(t)$ be the Elias gamma code of $t\ge 1$, of $2\lfloor\log t\rfloor+1$ bits. Fix an integer $\Phi\ge 1$; Appendix~\ref{app:bits} sets $\Phi=O(k\log^\ast n)$. The \emph{grain} of $y$ is $\mathrm{gr}(y)=2^{\epsilon(y)}$ with $\epsilon(y)=\max\{0,\lfloor\ell-4\log(\ell+2)-\log\Phi\rfloor\}$. Since $x-4\log(x+2)$ is negative for $x\le 4$ and increasing for $x\ge 4$, the grain is nondecreasing in $y$. Moreover
\[
\mathrm{gr}(y)\le\frac{y}{\Phi\lambda(y)^4}\quad\text{if }\mathrm{gr}(y)>1,\qquad\text{and}\qquad y<4\Phi\lambda(y)^4\,\mathrm{gr}(y)\quad\text{always.}
\]
The \emph{rounding} of $y$ is $\nu(y)=\mathrm{gr}(y)\lceil y/\mathrm{gr}(y)\rceil$, so that $y\le\nu(y)<y+\mathrm{gr}(y)$. Its \emph{code} consists of $\gamma(\ell+1)$ and the number $\lceil y/\mathrm{gr}(y)\rceil\le 4\Phi\lambda(y)^4$ in $\lceil\log(4\Phi\lambda(y)^4+1)\rceil$ bits; it determines $\nu(y)$ and $\mathrm{gr}(y)$ and has $O(\log\lambda(y)+\log\Phi)$ bits.

A \emph{container} is a set of consecutive positions. Every container $C$ of the construction has a \emph{raw length} $y_C$, a \emph{length} $|C|=\nu(y_C)$ and a grain $g_C=\mathrm{gr}(y_C)$, which divides $|C|$, and it is \emph{aligned}: its first position $\mathrm{st}(C)$ is a multiple of $g_C$. Its \emph{code} is the code of $y_C$.

\begin{lemma}[residues]\label{lem:residue}
Let $C$ be an aligned container and $q\in C$. Then $\mathrm{st}(C)=g_C(\lfloor q/g_C\rfloor-j)$ for an integer $0\le j<4\Phi\lambda(y_C)^4$. Hence $\mathrm{st}(C)$ is determined by $q$, the code of $C$ and $O(\log\lambda(y_C)+\log\Phi)$ further bits.
\end{lemma}
\begin{proof}
Both $\mathrm{st}(C)$ and $g_C\lfloor q/g_C\rfloor$ are multiples of $g_C$ in the range of $C$, which has $|C|/g_C=\lceil y_C/\mathrm{gr}(y_C)\rceil\le 4\Phi\lambda(y_C)^4$ multiples of $g_C$.
\end{proof}

\section{Proof of Proposition~\ref{prop:contig}}\label{app:contig}

\begin{proof}[Proof of Proposition~\ref{prop:contig}]
Let $m=2R+1$. A unit $U_0$ is a vertex $c$ of color $0$ with a child $b$ of color $1$, which has a child $w$ of color $3$. For $t\ge 1$, a unit $U_t$ with root $c$ of color $0$ consists of $c$, $b$ and $w$ as before, together with $m$ children $a_i$ of $c$ of color $1$; every $a_i$ has a child $u_i$ of color $2$, and every $u_i$ has two children, $y_i$ of color $1$ with a child $x_i$ of color $3$, and the root $c_i$, of color $0$, of a copy of $U_{t-1}$. Then $g_0(w)=g_0(x_i)=c$, and $|U_t|=3+m(4+|U_{t-1}|)$, so $|U_t|+8\le m(|U_{t-1}|+8)$ and $|U_d|\le 11m^d$.

We show by induction on $t$ that, for a unit with root $c$ at level $t$, some position in the range of $T_c$ lies in at least $t+1$ intervals of color $0$ owned by vertices of $T_c$. For $t=0$, the position of $w$ lies in an interval of $c$. For $t\ge 1$, the at most $R$ intervals of $c$ have at most $2R$ endpoints, and the ranges of the subtrees $T_{u_i}$ are disjoint, so the range $Q$ of some $T_{u_i}$ contains no endpoint, and every interval of $c$ contains $Q$ or misses it. The range $Q$ contains $x_i$, which lies in an interval $I$ of $c$ by correctness, so $Q\subseteq I$. By induction, some position in the range of $T_{c_i}\subseteq Q$ lies in $t$ intervals of color $0$ owned by vertices of $T_{c_i}$, and it also lies in $I$.
\end{proof}

\section{Properties of the Layout}\label{app:layout}

\begin{figure}[t]
\centering
\begin{tikzpicture}[scale=0.98]
\draw[->,thick,gray] (0,0)--(13.6,0);
\draw[decorate,decoration={brace,amplitude=5pt},thick] (0.4,0.85)--(6.8,0.85) node[midway,above=6pt] {\small $J$: a kept block of color $c$, owned by $o(J)$};
\node[cg,label=below:{\small $v$}] (u) at (1.4,0) {};
\node[cw,label=below:{\small $v'$}] at (3.7,0) {};
\draw[thick] (3.95,-0.05) rectangle (5.6,0.4); \node at (4.78,0.17) {\scriptsize $B(v')$};
\node[above] at (4.78,0.4) {\scriptsize light: kept};
\draw[thick,blue!70!black] (7.3,-0.15) rectangle (13.0,0.62);
\node[above,blue!70!black] at (10.15,0.62) {\small wrap of $z=g_{c'}(v)$, color $c'$};
\draw[thick,fill=gray!10] (7.6,-0.03) rectangle (10.3,0.45); \node at (8.95,0.21) {\scriptsize $B(v)$, a moved block};
\draw[thick,fill=gray!10] (10.6,-0.03) rectangle (12.7,0.45); \node at (11.65,0.21) {\scriptsize another moved block};
\draw[decorate,decoration={brace,amplitude=5pt,mirror},thick] (7.2,-0.35)--(13.1,-0.35) node[midway,below=6pt] {\small area of $J$};
\draw[->,dashed,thick,red!70!black] (1.4,-0.45) to[out=-35,in=-150] (8.4,-0.15);
\node[red!70!black] at (4.9,-2.05) {\scriptsize heavy: $\mass(B(v))\ge\tau(\sigma(J))$};
\end{tikzpicture}
\caption{The escape rule. Both $v$ and $v'$ lie in $J$, a kept block of color $c$, and have color $c$. The block of $v'$ is light compared with $J$ and is kept. The block of $v$ is heavy, so it becomes a moved block in the area of $J$, right after $J$. There it no longer lies in the intervals of color $c'$ that contained $v$, so it is wrapped in an interval of its upper neighbor $g_{c'}(v)$.}\label{fig:escape}
\end{figure}
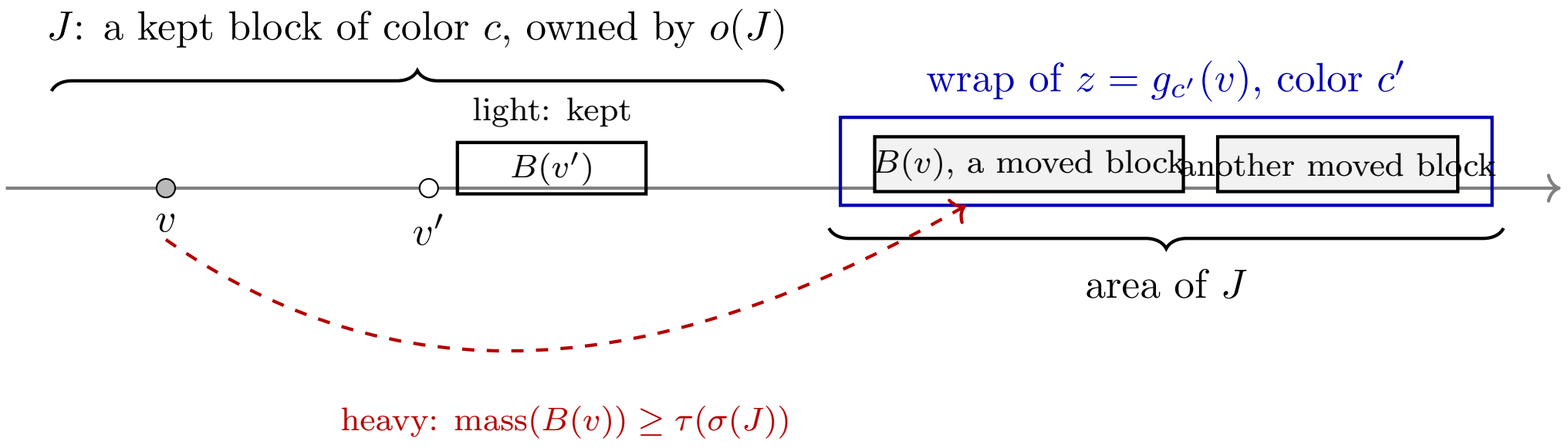

\begin{figure}[t]
\centering
\begin{tikzpicture}[scale=1]
\def\h{0.45}
\foreach \i/\a/\b/\lab in {1/0/1.6/{$z_1$},2/1.8/3.4/{$z_1z_2$},3/3.6/5.2/{$z_1z_2$},4/5.4/7.0/{$z_1z_3$},5/7.6/9.2/{$z_4$}}{
  \draw[thick,fill=gray!10] (\a,0) rectangle (\b,\h);
  \node at ({(\a+\b)/2},\h/2) {\scriptsize $Y_{\i}$};
  \node[below] at ({(\a+\b)/2},-0.05) {\scriptsize \lab};
}
\draw[thick,blue!70!black] (1.7,-0.55) rectangle (5.3,\h+0.25); \node[blue!70!black,above] at (3.5,\h+0.25) {\scriptsize wrap of $z_2$};
\draw[thick,green!50!black] (5.3,-0.6) rectangle (7.1,\h+0.25); \node[green!50!black,above] at (6.2,\h+0.25) {\scriptsize wrap of $z_3$};
\draw[thick] (-0.15,-0.75) rectangle (7.25,\h+0.75); \node[above] at (3.55,\h+0.75) {\small wrap of $z_1$};
\draw[thick,red!60!black] (7.45,-0.75) rectangle (9.35,\h+0.75); \node[red!60!black,above] at (8.4,\h+0.75) {\small wrap of $z_4$};
\node[left] at (-0.3,-0.25) {\scriptsize needs:};
\end{tikzpicture}
\caption{Five moved blocks with a common context and the vertices they need, shown below each moved block (all with the same sources). Moved blocks that agree on a prefix share the wraps of that prefix, which nest like a trie.}\label{fig:trie}
\end{figure}
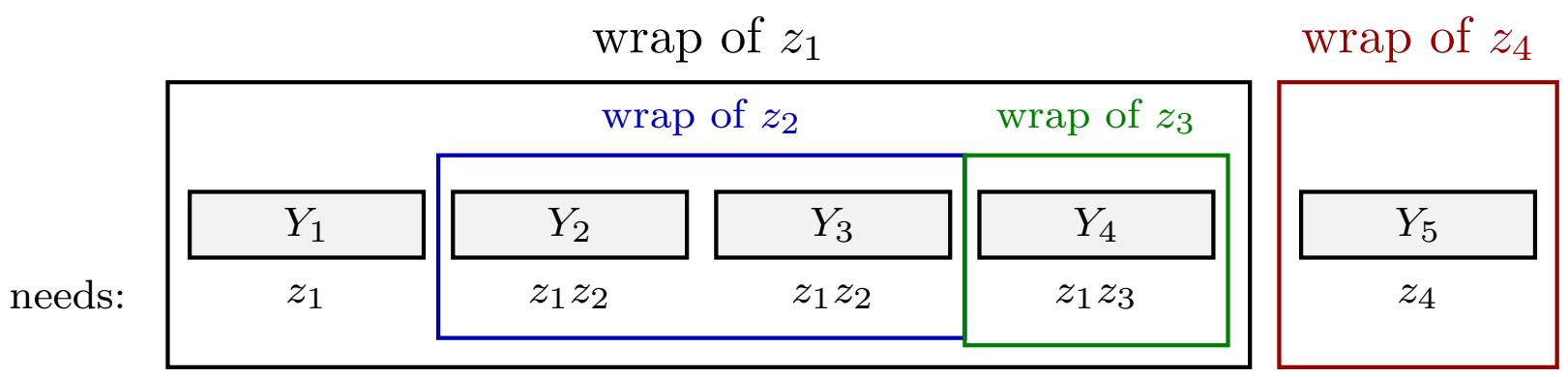

We use the following terms. For a moved block $B(v)$ with home $J$, the node $P$ of (E3) is its \emph{context}; the moved block is a \emph{primary} if $J$ is kept and a \emph{secondary} otherwise, and the wraps $N_1,\dots,N_m$ of (E5) are its \emph{chain}. We write $\sigma_A=s(o(J_A))$. Table~\ref{tab:notation} recalls the notation, and Figures~\ref{fig:escape} and~\ref{fig:trie} illustrate moved blocks and shared wraps.

\begin{table}[t]
\centering\small
\begin{tabular}{ll}
\hline
$g_c(x)$ & nearest proper ancestor of $x$ of color $c$\\
$B(v)$ & the children of $v$: an interval of color $\col(v)$ owned by $v$, of mass $s(v)-1$\\
$J$, $P$ & home and context of a moved block: it leaves $J$ and hangs below $P$\\
$\sigma(L)$ & subtree size of the owner of the kept block that $L$ belongs to\\
$\tau$ & the threshold; a block is moved iff its mass is at least $\tau(\sigma(J))$\\
$S_i$ & source of the $i$-th needed vertex: its interval around the leaf of the owner\\
$\alpha_c(x)$ & number of intervals of color $c$ containing $x$\\
\hline
\end{tabular}
\caption{Notation.}\label{tab:notation}
\end{table}

\paragraph{Areas and units.} The moved blocks and wraps of an area are its \emph{units}. The units of $A$ form a forest below the node $A$, in which every wrap is the parent of the units created through it; a primary hangs below the area itself, and a secondary below the parent of its home, a unit of the same area. The moved blocks with a common context share their wraps as far as their pairs $(z_i,S_i)$ agree, so these wraps nest like a trie (Figure~\ref{fig:trie}). A moved block moved out of $J$ is written right after $J$ in the area of $J$ if $J$ is kept, and next to $J$ otherwise. In both cases it lies outside $J$, as the escape rule requires, and its wraps restore the intervals of other colors that it left.

We start with the following simple observations on the layout tree.

\begin{lemma}\label{lem:facts}
\begin{enumerate}[label=(\alph*)]
\item The interval ancestors of the context $P$ of a moved block $B(v)$ are interval ancestors of the leaf $v$; for a primary they are the strict interval ancestors of its home.
\item Every leaf below an interval $L$ is a proper descendant of $o(L)$.
\item The units of an area $A$ lie below the node $A$, and none of them lies below a moved block of $A$. Every leaf below a wrap lies below a moved block below that wrap.
\end{enumerate}
\end{lemma}
\begin{proof}
(a) For a primary, $P$ is a child of the parent of the home $J$, so its interval ancestors are the strict interval ancestors of $J$, which is itself an interval ancestor of the leaf $v$. For a secondary, $P$ is the parent of $J$. (b) By induction over the attachments. A leaf $w$ is attached below $B(v)$ with $v=p(w)$. If $B(v)$ is kept, its strict interval ancestors are the interval ancestors of the leaf $v$; if it is a moved block, they are its chain, whose owners are proper ancestors of $v$, and the interval ancestors of $P$, which are interval ancestors of the leaf $v$ by (a). By induction $v$, and hence $w$, is a proper descendant of all their owners. (c) Units of $A$ are attached below $A$ or below a wrap of $A$, and never below a moved block of $A$. The children of a wrap are units, and an area is a child of a block.
\end{proof}

\begin{lemma}[sources]\label{lem:sources}
Let $B(v)$ be a moved block with home $J$, context $P$, needed vertices $z_1,\dots,z_m$ and sources $S_1,\dots,S_m$. Then
\begin{enumerate}[label=(\alph*)]
\item $S_i$ is owned by $z_i$ and lies strictly below $J$, and every interval ancestor of $P$ is a strict interval ancestor of $S_i$;
\item for $L'$ strictly below $L$, both intervals, $o(L')$ is a proper descendant of $o(L)$;
\item $S_1\supsetneq S_2\supsetneq\dots\supsetneq S_m$, and $S_i$ is the nearest interval of color $\col(z_i)$ strictly above $S_{i+1}$.
\end{enumerate}
\end{lemma}
\begin{proof}
(a) By Theorem~\ref{thm:exact} at $v$, $S_i$ is owned by $g_{\col(z_i)}(v)=z_i$. If $S_i$ were an interval ancestor of $P$, then by Lemma~\ref{lem:facts}(a) it would be the nearest interval ancestor of $P$ of its color, and $z_i$ would not be needed. For a secondary, the interval ancestors of the leaf $v$ that are not interval ancestors of $P$ lie below the child $J$ of $P$; for a primary, the interval ancestors of $P$ are the strict interval ancestors of $J$ by Lemma~\ref{lem:facts}(a). In both cases $S_i$ lies below $J$, hence strictly below every interval ancestor of $P$, and $S_i\ne J$ since $\col(z_i)\ne\col(v)$.

(b) By induction over the creation of $L'$. If $L'$ is a kept block, its strict interval ancestors are those of its owner's leaf, and Lemma~\ref{lem:facts}(b) applies. If $L'$ is a moved block, they are its chain and the interval ancestors of its context, and Lemma~\ref{lem:facts}(a,b) applies. If $L'$ is the wrap $N_i$ of a moved block as in (E5), its strict interval ancestors are $N_1,\dots,N_{i-1}$, owned by proper ancestors of $z_i$, and the interval ancestors $L$ of $P$. Every such $L$ lies strictly above $S_i$ by (a), and $S_i$ existed before $N_i$, so $z_i=o(S_i)$ is a proper descendant of $o(L)$ by induction.

(c) The sources are interval ancestors of the leaf $v$, hence pairwise comparable. If $S_i$ were below $S_{i+1}$, then $z_i$ would be a descendant of $z_{i+1}$ by (b), but it is a proper ancestor. An interval of color $\col(z_i)$ strictly between $S_{i+1}$ and $S_i$ would be nearer to the leaf $v$ than $S_i$.
\end{proof}

In the following lemma we prove that the block of a needed vertex lies inside every interval of its color that the moved block keeps. All quantitative statements below rely on it.

\begin{lemma}[needed blocks]\label{lem:needed}
Let $W$ be a wrap of $w$. Then $B(w)$ lies strictly below every interval of color $\col(w)$ that is a strict interval ancestor of $W$.
\end{lemma}
\begin{proof}
By induction over the creation of $W$, for a moved block with context $P$. Between $W$ and $P$ there are only wraps of the chain, of colors other than $\col(w)$, so every strict interval ancestor $L$ of $W$ of color $\col(w)$ is an interval ancestor of $P$. The source $S$ of $W$ is owned by $w$ and lies strictly below $L$ by Lemma~\ref{lem:sources}(a). If $S=B(w)$ we are done. Otherwise $S$ is a wrap of $w$ created earlier, $L$ is a strict interval ancestor of $S$ of color $\col(w)$, and induction applies.
\end{proof}

\begin{lemma}\label{lem:sigma}
\begin{enumerate}[label=(\alph*)]
\item If an interval $L'$ lies below an interval $L$, then $\sigma(L')\le\sigma(L)$.
\item Every moved block of an area $A$ has mass at least $\tau(\sigma_A)$, where $\sigma_A=s(o(J_A))$, and its owner is a proper descendant of $o(J_A)$.
\item \emph{(Lightness.)} Let $R$ be a block of color $c$ and $K$ the nearest strict interval ancestor of $R$ of color $c$. Then $\mass(R)<\tau(\sigma(K))$.
\item Two wraps of the same color, one below the other, belong to different areas.
\end{enumerate}
\end{lemma}
\begin{proof}
(b) A moved block of $A$ is heavy for its home, which is $J_A$ or a unit of $A$, and both have $\sigma=\sigma_A$. Its owner's leaf lies below $J_A$, or below a moved block of $A$ with a shallower owner (Lemma~\ref{lem:facts}(c)), and induction on the depth of the owner gives the second claim.

(a) Let $b(L)$ be $L$ if $L$ is kept and $J_A$ if $L$ belongs to $A$, so that $\sigma(L)=s(o(b(L)))$. If $L$ is kept, then $b(L')$ lies below $L$: this is clear if $L'$ is kept, and if $L'$ belongs to an area $A'$, then $L$ is a strict ancestor of the node $A'$, since otherwise $L$, and with it $L'$, would lie below a moved block of $A'$, contrary to Lemma~\ref{lem:facts}(c); hence $L$ is also a strict ancestor of the sibling $J_{A'}$ of that node. Lemma~\ref{lem:sources}(b) gives the claim. If $L$ belongs to $A$, then $L'$ belongs to $A$, with equal $\sigma$, or lies below a moved block $Y$ of $A$ by Lemma~\ref{lem:facts}(c), and then $s(o(b(L')))<s(o(Y))<\sigma_A$ by (b).

(c) By induction on the depth of $o(R)$. If $R$ is kept, $K$ is the interval $J$ of (E2). Let $R=B(v)$ be a moved block with home $J$ and context $P$. Its chain has other colors, so $K$ is the nearest interval ancestor of $P$ of color $c$, which is also the nearest strict interval ancestor of $J$ of color $c$. If $J$ is a block, $v$ is a proper descendant of $o(J)$ and $\mass(R)<\mass(J)<\tau(\sigma(K))$ by induction. If $J$ is a wrap of $w$, then $v$ lies below a moved block below $J$, so $v$ is a proper descendant of $w$ and $\mass(R)<\mass(B(w))$; by Lemma~\ref{lem:needed}, $B(w)$ lies strictly below $K$, so by induction and (a), $\mass(B(w))<\tau(\sigma(K))$.

(d) Let $W'$, a wrap of $a'$, lie strictly below $W$, of the same color, both in $A$, and let $W'$ be created for the moved block $Z$ of $A$. By Lemma~\ref{lem:needed}, (c) and (a), $\mass(B(a'))<\tau(\sigma(W))=\tau(\sigma_A)$, while $\mass(Z)<\mass(B(a'))$ since $Z$ needs $a'$, and $\mass(Z)\ge\tau(\sigma_A)$ by (b).
\end{proof}

\section{Proofs for Section~\ref{sec:bounds}}\label{app:depth}

\begin{figure}[t]
\centering
\begin{tikzpicture}[scale=0.85]
\def\q{6.2}
\foreach \i/\a/\b/\y in {1/0.2/12.2/1.6,2/1.2/11.2/1.2,3/2.2/10.2/0.8,4/3.2/9.2/0.4}{
  \draw[thick] (\a,-0.05)--(\a,\y)--(\b,\y)--(\b,-0.05);
  \node[above] at ({\a+0.9},\y) {\scriptsize $R_{\i}$};
}
\draw[->,thick,gray] (0,0)--(12.6,0);
\fill (\q,0) circle (2pt); \node[below] at (\q,-0.05) {\scriptsize $q$};
\node[right,align=left] at (12.8,0.9) {\small $\mass(R_4)<\tau(\sigma(R_3))$\\[2pt] \small $\sigma(R_3)\le\mass(R_2)$};
\end{tikzpicture}
\caption{Nested blocks of one color around a point $q$. Every block is light compared with the interval of its color around it (Lemma~\ref{lem:sigma}), and $\sigma$ of a block is at most the mass of the block around it, since the kept block it came from lies inside. So the masses drop by $\tau$ every two levels.}\label{fig:depth}
\end{figure}
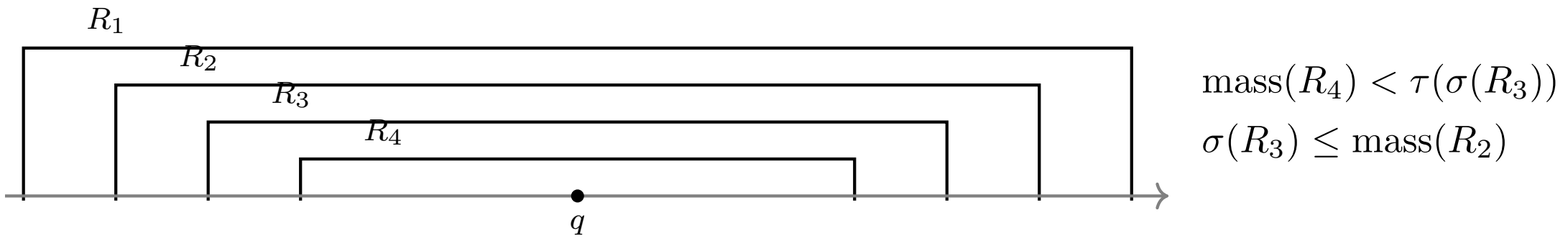

\begin{proof}[Proof of Theorem~\ref{thm:depth}]
We first show that if a block $R'$ lies strictly below a block $R$, then $\sigma(R')\le\mass(R)$. If $R'$ is kept, $o(R')$ is a proper descendant of $o(R)$ by Lemma~\ref{lem:sources}(b). If $R'$ is a moved block of $A$, then $R$ does not lie below the node $A$, since otherwise $R'$ would lie below a moved block of $A$, contrary to Lemma~\ref{lem:facts}(c); so $R$ is a strict ancestor of the node $A$ and of its sibling $J_A$, and $o(J_A)$ is a proper descendant of $o(R)$.

Fix a leaf and a color $c$, and let $R_1\supsetneq R_2\supsetneq\cdots$ be the blocks of color $c$ containing it, with masses $x_1>x_2>\cdots$. The nearest strict interval ancestor of $R_{i+1}$ of color $c$ lies below $R_i$, so Lemma~\ref{lem:sigma}(c,a) and the claim give $x_{i+1}<\tau(\sigma(R_i))\le\tau(x_{i-1})$ (Figure~\ref{fig:depth}). Hence $x_{2j+1}<\tau^j(n)$, and once this is at most $m_\tau$ the masses still decrease strictly; this bounds the number of blocks.

Let $W_1\supsetneq W_2\supsetneq\cdots$ be the wraps of color $c$ containing the leaf, owned by $a_1,a_2,\dots$, with $x_j=\mass(B(a_j))$; the $a_j$ descend strictly. By Lemma~\ref{lem:needed}, $B(a_{j+1})$ lies strictly below $W_j$, so $x_{j+1}<\tau(\sigma(W_j))$ by Lemma~\ref{lem:sigma}(c,a). By Lemma~\ref{lem:sigma}(d), $W_j$ belongs to an area $A_j$ different from that of $W_{j-1}$, so $W_{j-1}$ is a strict ancestor of the node $A_j$, and $J_{A_j}$ lies below a moved block $Y$ below $W_{j-1}$. Then $\sigma(W_j)=s(o(J_{A_j}))<s(o(Y))<s(a_{j-1})$, and $x_{j+1}<\tau(x_{j-1})$.

\end{proof}

\begin{proof}[Proof of Lemma~\ref{lem:iter}]
Let $\ell_j=\log^{(j)}n$ and $s_j=\tau_\beta^j(n)$. We claim $s_j\le(2\beta\ell_j)^\beta$ as long as $\ell_j\ge\log(4\beta)$. For $j=1$, $s_1\le(\log n+1)^\beta$. If the claim holds for $j$, then $s_{j+1}\le(\log s_j+1)^\beta\le(\beta\log(2\beta)+\beta\ell_{j+1}+1)^\beta\le(2\beta\ell_{j+1})^\beta$ when $\ell_{j+1}\ge\log(4\beta)$. Let $j^\ast$ be the last $j$ with $\ell_j\ge\log(4\beta)$; then $j^\ast\le\log^\ast n$, $\ell_{j^\ast}<4\beta$ and $s_{j^\ast}\le(8\beta^2)^\beta$. From there $s\mapsto\lceil s/2\rceil$ reaches $1$ in at most $\beta\log(8\beta^2)+1$ steps.
\end{proof}

\subsection{Intervals per Vertex}\label{app:records}

We call the intervals of a vertex $z$ its \emph{records}.

\begin{lemma}\label{lem:persource}
Every record $S$ of $z$ is the source of at most $k2^{k}$ wraps of $z$.
\end{lemma}
\begin{proof}
Let $W$ be a wrap of $z$ with source $S$, created for the moved block $B(v)$ with home $J$, needed vertices $z_1,\dots,z_m$ and $z=z_i$. The leaf $v$ lies below $S$, and $J$ is the nearest interval ancestor of the leaf of color $\col(v)$, so by Lemma~\ref{lem:sources}(a), $J$ is the nearest interval of color $\col(v)\ne\col(z)$ strictly above $S$. It determines the context $P$. Going up from $S_i=S$, Lemma~\ref{lem:sources}(c) determines $S_{i-1},\dots,S_1$ and their owners from the colors of $z_1,\dots,z_{i-1}$, and these pairs determine $W$. There are at most $k$ choices for $\col(v)$ and $2^{k-1}$ for the set of colors.
\end{proof}

For a wrap $W$ of an area $A$ let $t(W)$ be the number of wraps of $A$ that are ancestors of $W$, $W$ included. By Lemma~\ref{lem:sigma}(d) the wraps of $A$ on a path have distinct colors, so $t(W)\le k+1$.

\begin{lemma}\label{lem:samearea}
If the source $S$ of a wrap $W$ of $z$ is a wrap of $z$ of the same area, then $t(W)<t(S)$.
\end{lemma}
\begin{proof}
Let $W$ be created for the moved block $Z$ with context $P$, home $J$, and $z=z_i$. The source $S$ lies strictly below $J$, which is a child of $P$, so $Z$ is a secondary, and since no unit of the area lies below a moved block of it (Lemma~\ref{lem:facts}(c)), $J$ is a wrap and every interval between $J$ and $S$ is a wrap of the area. These include $J$ and, by Lemma~\ref{lem:sources}, the sources $S_1\supsetneq\dots\supsetneq S_{i-1}$ of $Z$, which lie strictly between $J$ and $S$. Hence $t(S)\ge t(P)+i+1$, where $t(P)=0$ if $P$ is the area, while $t(W)=t(P)+i$.
\end{proof}

\begin{lemma}[escape chains]\label{lem:chains}
Let $W_1,W_2,\dots,W_r$ be wraps of $z$, each the source of the next. Let $t_1<t_2<\dots$ be the indices $t$ for which $W_{t+1}$ belongs to a different area than $W_t$, and let $K_j$ be the home of $W_{t_j+1}$. Then $K_1\subsetneq K_2\subsetneq\cdots$, and these homes have distinct colors, all different from $\col(z)$. Hence there are at most $k$ such indices.
\end{lemma}
\begin{proof}
Let $A_t$ be the area of $W_t$. By Lemma~\ref{lem:sources}(a) the home $K$ of $W_{t+1}$ contains $W_t$ strictly. If $K$ were a unit of $A_t$, then $W_{t+1}$ would belong to $A_t$. So for $t=t_j$, $K_j$ is a strict ancestor of the node $A_t$. Every later wrap of the sequence up to index $t_{j+1}$ lies in the area $A_{t+1}$, which is the area of $K_j$ if $K_j$ is kept and the area containing $K_j$ otherwise; in both cases $K_{j+1}$, a strict ancestor of that area node, contains $K_j$ strictly. The color of $K_j$ is that of the moved block for which $W_{t_j+1}$ was created, which needs $z$, so it differs from $\col(z)$.

Suppose $K_a\subsetneq K_b$ have the same color $d$, and let $Z$ be the moved block for which $W_{t_b+1}$ was created; it has home $K_b$, needs $z$, and $\tau(\sigma(K_b))\le\mass(Z)<\mass(B(z))$. The wrap $W_{t_a}$ of $z$ lies strictly below $K_a$. If $K_a=B(y)$ is a block, then $z$ is a proper descendant of $y$ by Lemma~\ref{lem:sources}(b), and by Lemma~\ref{lem:sigma}, $\mass(B(z))<\mass(B(y))<\tau(\sigma(K_b))$, since the nearest strict interval ancestor of $K_a$ of color $d$ lies below $K_b$. If $K_a$ is a wrap of $y$, then again $\mass(B(z))<\mass(B(y))$, and $B(y)$ lies strictly below $K_b$ by Lemma~\ref{lem:needed}, so $\mass(B(y))<\tau(\sigma(K_b))$. Both contradict the first inequality.
\end{proof}

\begin{proof}[Proof of Theorem~\ref{thm:records}]
Give $B(z)$ generation $0$, and a wrap the generation of its source plus one. Along a sequence of wraps of $z$, each the source of the next, the area changes at most $k$ times by Lemma~\ref{lem:chains}, and between changes $t$ strictly decreases by Lemma~\ref{lem:samearea}, so the sequence has at most $(k+1)^2$ wraps. By Lemma~\ref{lem:persource} the number of records of generation $g$ is at most $(k2^k)^g$.
\end{proof}

\begin{proof}[Proof of Lemma~\ref{lem:depths}]
Let $W$ be created for the moved block $Z$ with context $P$ and home $J$. The intervals of color $\col(z)$ containing $W$ are $W$ and the interval ancestors of $P$, which are the strict interval ancestors of $J$ (Lemma~\ref{lem:facts}(a) and the definition of $P$). Those containing $S$ are the same strict interval ancestors of $J$, the intervals strictly between $J$ and $S$, and $S$; the home $J$ has another color. So $\delta$ is the number of intervals of color $\col(z)$ strictly between $J$ and $S$, and they contain the leaf of $o(Z)$. Suppose there are two, $L_1\subsetneq L_2$. If $L_1=B(y)$, then $\mass(Z)<\mass(B(y))<\tau(\sigma(L_2))\le\tau(\sigma(J))$ by Lemma~\ref{lem:sigma}. If $L_1$ is a wrap of $y$, then the leaf of $o(Z)$ lies below a moved block below $L_1$, so again $\mass(Z)<\mass(B(y))$, and $B(y)$ lies strictly below $L_2$ by Lemma~\ref{lem:needed}. Both contradict $\mass(Z)\ge\tau(\sigma(J))$.
\end{proof}

\section{The Identifier Layer}\label{app:bits}

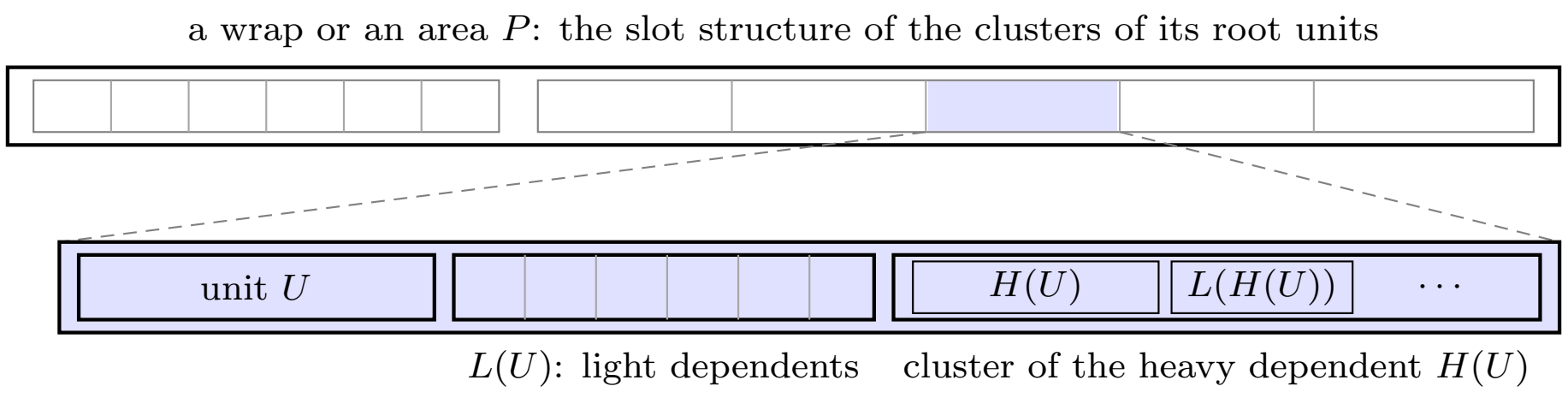
\begin{figure}[t]
\centering
\begin{tikzpicture}[x=1cm,y=1cm]
\draw[thick] (0,0) rectangle (12,0.6);
\node[font=\scriptsize,anchor=south] at (6,0.62) {a wrap or an area $P$: the slot structure of the clusters of its root units};
\draw[gray] (0.2,0.1) rectangle (3.8,0.5);
\foreach \t in {0.8,1.4,...,3.3}{ \draw[gray!70] (\t,0.1)--(\t,0.5); }
\draw[gray] (4.1,0.1) rectangle (11.8,0.5);
\foreach \t in {5.6,7.1,8.6,10.1}{ \draw[gray!70] (\t,0.1)--(\t,0.5); }
\fill[blue!12] (7.12,0.11) rectangle (8.58,0.49);
\draw[gray,densely dashed] (7.1,0.1)--(0.4,-0.75); \draw[gray,densely dashed] (8.6,0.1)--(12,-0.75);
\draw[thick,fill=blue!12] (0.4,-1.45) rectangle (12,-0.75);
\draw[thick] (0.55,-1.35) rectangle (3.3,-0.85); \node[font=\scriptsize] at (1.92,-1.1) {unit $U$};
\draw[thick] (3.45,-1.35) rectangle (6.7,-0.85);
\foreach \t in {4.0,4.55,...,6.3}{ \draw[gray!70] (\t,-1.35)--(\t,-0.85); }
\node[font=\scriptsize,anchor=north] at (5.07,-1.45) {$L(U)$: light dependents};
\draw[thick] (6.85,-1.35) rectangle (11.85,-0.85);
\draw (7.0,-1.3) rectangle (8.9,-0.9); \node[font=\scriptsize] at (7.95,-1.1) {$H(U)$};
\draw (9.0,-1.3) rectangle (10.4,-0.9); \node[font=\scriptsize] at (9.7,-1.1) {$L(H(U))$};
\node[font=\scriptsize] at (11.1,-1.1) {$\cdots$};
\node[font=\scriptsize,anchor=north] at (9.35,-1.45) {cluster of the heavy dependent $H(U)$};
\end{tikzpicture}
\caption{Placement of units inside a wrap or an area $P$. The clusters of the units of $P$ that depend on no other unit are placed in the slots of a slot structure, grouped into sub-areas by their rounded length (here, one sub-area with short slots and one with long slots); a cluster occupies a slot determined by the position of its key and a short choice. Inside a cluster, a unit $U$ is followed by the slot structure $L(U)$ of the clusters of its light dependents, and then by the cluster of its heavy dependent $H(U)$.}\label{fig:cluster}
\end{figure}

In this appendix we fix the order of the layout tree, insert gaps, and prove Lemmas~\ref{lem:locate} and~\ref{lem:size}. The owner of a unit $U$ knows the position of the \emph{key} of $U$: its source if $U$ is a wrap, and its own position if $U$ is a moved block. We place $U$ so that it can be found from its key with few bits. By Theorem~\ref{thm:place} below, the key lies inside the base of the parent of $U$, or inside a sibling of $U$ on which $U$ depends. We therefore place every unit right after the unit it depends on, together with everything that depends on it, as one \emph{cluster}. The clusters that depend on nothing are placed in their parent by a \emph{slot structure}, a hash table without collisions whose slot is found from the position of the key and a short choice (Figure~\ref{fig:cluster}). All lengths are rounded and all intervals are aligned, so that the start of an interval can be recovered from any position inside it and a few bits.

\subsection{Slot Graphs}

Slot graphs are hash functions with $d$ choices per key, as in $d$-ary cuckoo hashing~\cite{FPSS05}; since we need one function that works for every set of keys, we prove their existence by counting.

\begin{lemma}[slot graphs]\label{lem:slots}
For all integers $Q\ge m\ge 1$ and $0<\varepsilon\le 1$, let $M=m+\lceil\varepsilon m\rceil$ and $d=\lceil 2(\ln(Q/m)+6)/\varepsilon\rceil$. There is a function $h:[Q]\times[d]\to[M]$ such that for every set $S\subseteq[Q]$ of at most $m$ keys there is a choice $b:S\to[d]$ for which $q\mapsto h(q,b(q))$ is injective on $S$.
\end{lemma}
\begin{proof}
Choose every value $h(q,i)$ independently and uniformly in $[M]$. By Hall's theorem, a set $S$ fails only if some $S'\subseteq S$ with $|S'|=s\ge 2$ sends all its $ds$ values into a set of $s-1$ slots. The failure probability is therefore at most $\sum_{s=2}^m x_s$ with $x_s=\binom{Q}{s}\binom{M}{s-1}((s-1)/M)^{ds}$. If $s\le m/2$, then $\binom{Q}{s}\le(eQ/s)^s$ and $\binom{M}{s-1}\le(eM/s)^s$, and $M/s\ge 2$ gives $x_s\le(e^2(Q/M)(s/M)^{d-2})^s\le(e^2(Q/m)2^{-(d-2)})^s\le 4^{-s}$, since $d\ge 2\ln(Q/m)+12$. If $m/2<s\le m$, then $\binom{Q}{s}\le(2eQ/m)^s$, $\binom{M}{s-1}\le 2^M\le 2\cdot 16^s$ and $((s-1)/M)^{ds}\le(1+\varepsilon)^{-ds}\le e^{-d\varepsilon s/2}\le((m/Q)e^{-6})^s$, so $x_s\le 2(32e^{-5})^s\le 2\cdot0.22^s$. The sum is less than $1$.
\end{proof}

We fix $h_{Q,m,\varepsilon}$ as the lexicographically first such function. If $m\le 1/\varepsilon$ we use $M=m$, $d=m$ and $h(q,i)=i$ instead. In both cases a choice $b(q)$ takes $\log(\ln(Q/m)+6)+\log(1/\varepsilon)+O(1)$ bits.

\subsection{Keys}

The \emph{key} $\kappa(U)$ of a unit $U$ is its source if $U$ is a wrap, and the leaf of its owner if $U$ is a moved block. The \emph{base} of a node $P$ that is an area or a wrap is $\Base(P)=J_A$ if $P$ is the area $A$, and the source of $P$ if $P$ is a wrap.

\begin{theorem}[keys]\label{thm:place}
Let $U$ be a unit with parent $P$, created for a moved block with home $J$. Then $\kappa(U)$ is not below $U$. If $U$ is the first wrap of the chain of a secondary, or a secondary with an empty chain, then $\kappa(U)$ lies strictly below the sibling $J$ of $U$, and $o(U)$ is a proper descendant of $o(J)$; we call $U$ a \emph{dependent} of $J$. Otherwise $\kappa(U)$ lies strictly below $\Base(P)$, and $\Base(P)$ is $J$ or lies below $J$.
\end{theorem}
\begin{proof}
Let $U=N_i$ in the notation of (E5), or $U=B(v)$ with $i=m+1$. The key existed before $U$ was created, so it is not below $U$. If $i\ge 2$, the parent is $N_{i-1}$, whose base $S_{i-1}$ contains $S_i$ (or the leaf $v$) strictly by Lemma~\ref{lem:sources}(c), and $S_{i-1}$ lies below $J$. If $i=1$, the parent is the context $P$ and the key lies strictly below $J$ by Lemma~\ref{lem:sources}(a). For a primary $\Base(P)=J$. For a secondary $J$ is a child of $P$ other than $U$, and $o(U)$ is a proper descendant of $o(J)$ by Lemma~\ref{lem:sources}(b).
\end{proof}

In other words, the \emph{dependency} relation among the children of a node $P$ is a forest whose roots, the \emph{root units} of $P$, have their keys below $\Base(P)$. Every unit $U$ has a \emph{walk}: $U_0=U$, and $U_{t+1}$ is the parent of $U_t$ as long as $U_t$ is not a dependent and its parent is a wrap. The walk follows the chain of the moved block that created $U$ and stops at a unit $U_r$, $r\le k$, which is a root unit of an area $A$ with key below $J_A=J$, or a dependent of $J$. In both cases all units of the walk have home $J$, and their bases are $J$ or lie below $J$.

\subsection{Containers}

We now fix the order of $\cL$ and insert gaps. Every vertex $v$ gets an \emph{alignment} $z(v)\ge 0$, chosen in Section~\ref{sec:labels}, and $\id(v)$ will be a multiple of $2^{z(v)}$. An aligned container is placed at the least multiple of its grain at or after the current position. Raw lengths are computed bottom-up:
\begin{enumerate}[label=(C\arabic*)]
\item The \emph{cell} of a vertex $v$ consists of $2^{z(v)}$ positions, with $v$ at the multiple of $2^{z(v)}$ among them, then $B(v)$ if it is kept, at the least multiple of its grain after $v$, then the area of $B(v)$ if it exists. Its raw length is $2^{z(v)}+|B(v)|+g_{B(v)}+|\text{area}|$, with the terms for $B(v)$ and the area omitted if absent.
\item A block, kept or moved, consists of the cells of the children of its owner, in any fixed order; its raw length is their total.
\item For a unit $U$ with dependents, let $H(U)$ be the dependent whose cluster contains the most vertices, and call the others \emph{light}; this is the heavy-light decomposition of~\cite{ST83}, applied to the dependency forest. The \emph{cluster} of $U$ is $U$, then $L(U)$, then the cluster of $H(U)$, where $L(U)$ is the slot structure of the clusters of the light dependents of $U$, with base $U$. Its raw length is $|U|+|L(U)|+g_{L(U)}+(\text{raw length of the cluster of }H(U))+g_{H(U)}$. The cluster of a unit without dependents is the unit itself.
\item A \emph{slot structure} with base $B$ for a set of clusters groups them into \emph{classes} by the code of their raw lengths. A class of $m$ clusters of rounded length $c$ gets a \emph{sub-area} of $M$ slots of length $c$, with $\varepsilon=(\Phi\lambda(c)^4)^{-1}$, where $M=m$ if $m\le 1/\varepsilon$, and otherwise $M=\hat m+\lceil\varepsilon\hat m\rceil$, where $\hat m$ is $m$ rounded up to a multiple of $2^{\lfloor\log(\varepsilon m)\rfloor}$; then $M\le(1+4\varepsilon)m$, and $\hat m$ is coded by $\lfloor\log m\rfloor$ and $\hat m/2^{\lfloor\log(\varepsilon m)\rfloor}\le 2/\varepsilon+1$. The sub-areas are aligned and placed in order of increasing length. The cluster of a unit $U$ occupies the slot $h(\kappa',b(U))$ of its class, with $h=h_{Q,\hat m,\varepsilon}$, $Q=4(k+1)(|B|+1)^2$, and $\kappa'\in[Q/2]$ encoding the color of $o(U)$, whether $\kappa(U)$ is a leaf, and the first and last positions of $\kappa(U)$ relative to $\mathrm{st}(B)$. Distinct units of one slot structure have distinct values $\kappa'$ (see below), so $m\le Q/2$ and $\hat m\le 2m\le Q$, as Lemma~\ref{lem:slots} requires.
\item A wrap $W$ consists of the slot structure with base $\Base(W)$ of the clusters of its root units. The area of $J$ consists of the slot structure with base $J$ of the clusters of its root units, and is placed right after $J$.
\end{enumerate}
Blocks and wraps have the length and grain of their raw length and are aligned; slots start at multiples of the grain of the class, which is at least the grain of the unit at the start of the cluster, since grains are nondecreasing. Two units of one slot structure have different keys $\kappa'$: distinct intervals of one color that are nested have different first or last positions, because a wrap of an area has no unit of its color below it in that area (Lemma~\ref{lem:sigma}(d) and induction) and every other interval starts after the first leaf of the block containing it. The positions are assigned top-down, starting with the cell of the root of $T$ at position $0$, and $N$ is the length of that cell.

\subsection{Locating Units}

For a home $J$, let $\mathrm{Reg}(J)$, the \emph{region} of $J$, be $J$ together with the area of $J$ if $J$ is kept, and $J$ together with $L(J)$ and $H(J)$ if $J$ is a unit. All units with home $J$ lie in $\mathrm{Reg}(J)$, since they lie in dependents of $J$ or in the area of $J$. Let $y(J)$ be the total raw length of the containers forming $\mathrm{Reg}(J)$. The set of vertices in $\mathrm{Reg}(J)$ does not depend on the alignments, since heavy dependents are chosen by numbers of vertices. Let $\hat y(J)=2^{10}\sum_{u}2^{z(u)}$, the sum over the vertices $u$ in $\mathrm{Reg}(J)$; Lemma~\ref{lem:size} below shows $y(J)\le\hat y(J)$.

\begin{proof}[Proof of Lemma~\ref{lem:locate}]
Let $U_0,\dots,U_r$ be the walk of $U$. Going up, $\kappa(U_t)$ lies in $\Base(U_{t+1})=\kappa(U_{t+1})$, an aligned container, so Lemma~\ref{lem:residue} gives its first position from the code of the base and $O(\log\lambda(y(J))+\log\Phi)$ bits, and the code gives its last position. At the top, $\kappa(U_r)$ lies in $J$, and the same gives the first and last positions of $J$. If $U_r$ is a root unit of the area of $J$, the area starts at the end of $J$; its sub-areas are found from their codes and their offsets, which are below $\#\text{classes}\cdot(4\Phi\lambda^4+1)$ in units of their grains since they are sorted by length, and the slot from the class code, $\hat m$, the key $\kappa'$ and $b(U_r)$. If $U_r$ is the heavy dependent of $J$, it starts at the next multiple of its grain after $L(J)$, which ends at a position given by the code of $L(J)$; if it is light, it is found in $L(J)$ like a root unit, with base $J$. Going down, the cluster of $U_{t}$ is found in the slot structure of $U_{t+1}$ in the same way, with base $\kappa(U_{t+1})$, whose position is known from the way up. The number of classes in a slot structure of total raw length $y$ is at most $\lambda(y)\cdot 4\Phi\lambda(y)^4$. All codes, offsets and choices refer to containers of raw length at most $y(J)$, and $\log\log Q=O(\log\lambda(y(J))+\log k)$.
\end{proof}

\subsection{The Identifier Space}

We prove the following form of Lemma~\ref{lem:size}: $N\le e^6(n+W)$, and every container $C$ formed by blocks, wraps, slot structures and clusters has raw length at most $e^6$ times the number of vertices in it plus the alignments $2^{z(u)}$ of the vertices $u$ in it.

\begin{proof}[Proof of Lemma~\ref{lem:size}]
Call a vertex together with its alignment positions an \emph{atom}. Rounding and aligning a container of raw length $y$ adds at most $2\,\mathrm{gr}(y)\le 2y/(\Phi\lambda(y)^4)$ positions, and nothing if $\mathrm{gr}(y)=1$; a sub-area multiplies the rounded lengths of its clusters by at most $1+4/(\Phi\lambda(c)^4)$ by the choice of $M$. Expanding the raw lengths top-down, every atom is inflated by the product of $1+6/\Phi$ over the blocks, wraps, slot structures, sub-areas and rounded clusters that contain it. These are its interval ancestors, at most $\Phi/8$ of them, at most three structures per area or wrap above it, and the sub-areas of the light clusters containing it. The light clusters at one node that contain the atom are nested, and the number of vertices in them at least halves from one to the next, since a light cluster has no more vertices than the heavy one. Their rounded lengths are at least these numbers, so the factors $1+6/(\Phi\lambda(c)^4)$ along such a chain multiply to at most $e^{1/\Phi}$. There are at most $\Phi/8$ areas and $\Phi/8$ wraps above an atom, so in total every atom is inflated by at most $(1+6/\Phi)^{7\Phi/8}e^{1/4}\le e^6$.
\end{proof}

\paragraph{Choosing the alignments.} The alignments determine the layout, and the layout determines how much data a vertex must absorb. Section~\ref{sec:labels} defines $z(v)$ by a rule that depends only on the numbers $\hat y(J)$ for the homes $J$ of the records of $v$ and is nondecreasing in each of them. Since the vertex sets $\mathrm{Reg}(J)$ do not depend on the alignments, the map $F$ that sends a vector of alignments to the vector defined by the rule is nondecreasing. Starting from $z\equiv 0$, the sequence $z,F(z),F(F(z)),\dots$ is therefore pointwise nondecreasing. It is also bounded: the rule gives $\max_vz(v)\le a\log(\max_uz(u)+\log n+12)+b$ for numbers $a,b$ that do not depend on the alignments, because $\lambda(\hat y(J))\le\log n+\max_uz(u)+12$. Hence the sequence reaches a fixed point $z=F(z)$, and we use the layout for this $z$. By Lemma~\ref{lem:size}, $y(J)\le 2e^6\sum_{u}2^{z(u)}\le\hat y(J)$ for every home $J$.

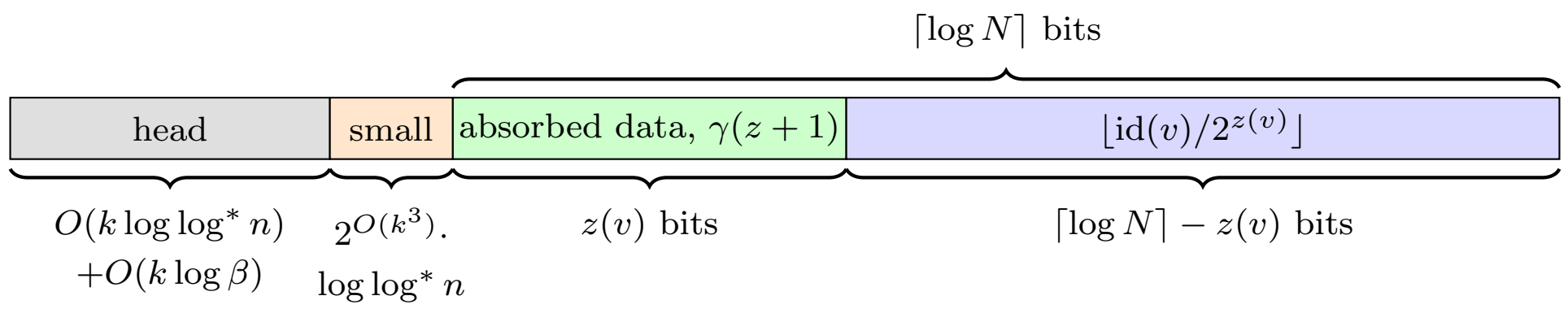
\begin{figure}[t]
\centering
\begin{tikzpicture}[scale=1]
\def\h{0.5}
\draw[fill=gray!25] (0,0) rectangle (2.6,\h); \node at (1.3,\h/2) {\scriptsize head};
\draw[fill=orange!20] (2.6,0) rectangle (3.6,\h); \node at (3.1,\h/2) {\scriptsize small};
\draw[fill=green!20] (3.6,0) rectangle (6.8,\h); \node at (5.2,\h/2) {\scriptsize absorbed data, $\gamma(z+1)$};
\draw[fill=blue!15] (6.8,0) rectangle (12.6,\h); \node at (9.7,\h/2) {\scriptsize $\lfloor\id(v)/2^{z(v)}\rfloor$};
\draw[decorate,decoration={brace,amplitude=4pt,mirror},thick] (0,-0.08)--(2.6,-0.08) node[midway,below=5pt,align=center] {\scriptsize $O(k\log\log^\ast n)$\\[-2pt] \scriptsize $+O(k\log\beta)$};
\draw[decorate,decoration={brace,amplitude=4pt,mirror},thick] (2.6,-0.08)--(3.6,-0.08) node[midway,below=5pt,align=center] {\scriptsize $2^{O(k^3)}\cdot$\\[-2pt]\scriptsize $\log\log^\ast n$};
\draw[decorate,decoration={brace,amplitude=4pt,mirror},thick] (3.6,-0.08)--(6.8,-0.08) node[midway,below=5pt] {\scriptsize $z(v)$ bits};
\draw[decorate,decoration={brace,amplitude=4pt,mirror},thick] (6.8,-0.08)--(12.6,-0.08) node[midway,below=5pt] {\scriptsize $\lceil\log N\rceil-z(v)$ bits};
\draw[decorate,decoration={brace,amplitude=4pt},thick] (3.6,\h+0.08)--(12.6,\h+0.08) node[midway,above=5pt] {\scriptsize $\lceil\log N\rceil$ bits};
\end{tikzpicture}
\caption{Composition of a label. The head holds the color, $k+1$ depths of at most $D_n=4\log^\ast n+O(\beta\log\beta)$ and the $k$ edge bits. The green part costs nothing beyond $\lceil\log N\rceil$: $\id(v)$ is a multiple of $2^{z(v)}$, and only the quotient is stored. It holds the data of the records whose homes have large regions. The data of the other records are stored explicitly (orange).}\label{fig:label}
\end{figure}

\section{Proof of Lemma~\ref{lem:waste}}\label{app:waste}

Let $c_1=4C_1(\rho_k+1)k$, let $K_0=2^{6C_1(\rho_k+1)k^3+5}$, let $\beta=4c_1+8$, and let $Z$ be a sufficiently large multiple of $c_1\log(c_1\Phi)+\log K_0+\beta\log\beta$. Then $c_1$, $\beta$ and $\log K_0$ are $2^{O(k^3)}$, and $Z=2^{O(k^3)}(1+\log\Phi)$. We prove
\[
W\le 4(k+2)^2K_0(10\Phi)^{c_1}\,n,
\]
which implies Lemma~\ref{lem:waste}.

\paragraph{Notation.} For a home $J$, a \emph{moved block with home $J$} is a moved block $B(x)$ whose home is $J$. By Theorem~\ref{thm:exact} applied to the color of $x$, $J$ is owned by $g_{\col(x)}(x)$, so no moved block with home $J$ has its owner below another one: the moved blocks with home $J$ have pairwise disjoint subtrees in $T$. Let $I(J)$ be their number, let $X_J$ be the number of vertices in $\mathrm{Reg}(J)$, and let $z_J$ be the largest alignment of a vertex in $\mathrm{Reg}(J)$. All units with home $J$ lie in $\mathrm{Reg}(J)$ and contain a vertex, so $I(J)\le X_J$. All vertices of $\mathrm{Reg}(J)$ are proper descendants of the owner of the kept block that $J$ belongs to, so $X_J<\sigma(J)$, and $\hat y(J)\le 2^{10}X_J2^{z_J}$, so that $\lambda(\hat y(J))\le\lambda(X_J)+z_J+11$.

\paragraph{Assignment.} Let $v$ be a vertex that absorbs data. A short computation shows $z(v)\le F_v+2\log(F_v+2)+3$, so $2^{z(v)}\le 8(F_v+2)^22^{F_v}\le 2^{2F_v+5}$. Since $v$ has at most $\rho_k$ records and $F_R$ is nondecreasing in $\hat y(J)$, this gives $2^{z(v)}\le K_0(\Phi\lambda(\hat y(J^\ast)))^{c_1}$, where $J^\ast$ is the home of the absorbed record $R^\ast$ of $v$ with the largest $\hat y(J^\ast)$. We \emph{assign} $v$ to $J^\ast$, and also to the moved block $R^\ast$, or to the moved block for which the wrap $R^\ast$ was created; if $R^\ast=B(v)$ is kept, we assign $v$ to $J^\ast=B(v)$ only. A moved block is assigned at most $k+1$ vertices, its owner and the owners of the wraps created for it, and the only other vertex that can be assigned to $J$ is the owner of $J$. Let $\mathrm{Ch}(J)$ be the sum of $2^{z(v)}$ over the vertices assigned to $J$. Then
\begin{equation}\label{eq:ch}
\mathrm{Ch}(J)\le K_0(k+2)(I(J)+1)(\Phi\lambda(\hat y(J)))^{c_1},\qquad\text{and }\lambda(\hat y(J))>4Z\text{ if }\mathrm{Ch}(J)>0.
\end{equation}
Every vertex that absorbs data contributes to exactly one $\mathrm{Ch}(J)$, and the vertices that absorb nothing have $2^{z(v)}=32$. We call a home $J$ with $\mathrm{Ch}(J)>0$ \emph{gap-dominated} if $z_J>2\lambda(X_J)+Z$, \emph{content-paid} if it is not gap-dominated and $\mathrm{Ch}(J)\le X_J/(\Phi\lambda(X_J)^3)$, and \emph{subtree-paid} otherwise. Claims~\ref{cl:gap},~\ref{cl:content} and~\ref{cl:subtree} below bound the total charge of the three kinds by $W/2$, $n$ and $(k+1)(k+2)K_0(10\Phi)^{c_1}n$, respectively, so
\[
W\le\tfrac12W+32n+n+(k+1)(k+2)K_0(10\Phi)^{c_1}n,
\]
and the bound on $W$ follows since $K_0\ge 64$. Both the gap-dominated and the content-paid homes are charged to vertices of their regions, so we first bound how many regions contain a vertex.

\begin{claim}[multiplicity of regions]\label{cl:mult}
Let $u$ be a vertex.
\begin{enumerate}[label=(\alph*)]
\item There are at most $3\Phi/8$ homes $J$ such that $u$ lies in $J$, in the area of $J$, or in the heavy dependent of $J$.
\item Every other home $J$ with $u\in\mathrm{Reg}(J)$ is a unit with $u\in L(J)$, and its parent is one of the at most $\Phi/4$ areas and wraps above $u$. If $J_1,\dots,J_N$ are the homes at one such node with $u\in L(J_i)$, ordered from the outside in, then $X_{J_i}\ge 2^{N-i-1}$, and hence $\lambda(X_{J_i})\ge N-i+1$.
\end{enumerate}
\end{claim}
\begin{proof}
(a) In each case some interval ancestor of $u$ determines $J$: the home $J$ itself, a unit of the area of $J$, or the heavy dependent of $J$. There are at most $\Phi/8$ interval ancestors of $u$.

(b) The clusters of the units of one node are nested or disjoint, and all of $J_1,\dots,J_N$ contain $u$ in their clusters. The cluster of $J_{i+1}$ lies in the cluster of $J_i$, but not in the cluster of $H(J_i)$, which is disjoint from $L(J_i)$. Hence it lies in the cluster $C_i$ of a light dependent of $J_i$. Since $C_i$ and the cluster of $H(J_i)$ are disjoint parts of the cluster of $J_i$, and the latter has at least as many vertices as $C_i$, the numbers of vertices of the clusters of $J_1,\dots,J_N$ at least halve from one to the next. Since $C_i\subseteq L(J_i)\subseteq\mathrm{Reg}(J_i)$, we get $X_{J_i}\ge|C_i|\ge 2^{N-i-1}$.
\end{proof}

\begin{claim}[gap-dominated homes]\label{cl:gap}
The total charge of the gap-dominated homes is at most $W/2$.
\end{claim}
\begin{proof}
Let $J$ be gap-dominated. Then $I(J)\le X_J<2^{\lambda(X_J)}<2^{(z_J-Z)/2}$ and $\lambda(\hat y(J))\le\lambda(X_J)+z_J+11\le 2z_J$, so by~\eqref{eq:ch},
\[
\mathrm{Ch}(J)\le 2^{z_J}\cdot 2^{-Z/2}\cdot 2K_0(k+2)(2z_J\Phi)^{c_1}2^{-z_J/2}\le\frac{2^{z_J}}{2(z_J+1)\Phi},
\]
where the last step uses $z_J>Z$ and that $Z$ is large. We charge $\mathrm{Ch}(J)$ to a vertex $u$ of $\mathrm{Reg}(J)$ with $z(u)=z_J$. A vertex $u$ is charged only by gap-dominated homes $J$ with $u\in\mathrm{Reg}(J)$, $z_J=z(u)$ and $X_J<2^{z(u)/2}$. By Claim~\ref{cl:mult}, there are at most $3\Phi/8$ such homes of the first kind, and at most $z(u)/2+1$ at each of the at most $\Phi/4$ areas and wraps above $u$, so at most $(z(u)+1)\Phi$ in total. Hence $u$ receives at most $2^{z(u)}/2$, and the total is at most $\sum_u2^{z(u)}/2=W/2$.
\end{proof}

Now let $J$ be a home with $\mathrm{Ch}(J)>0$ that is not gap-dominated. Then $\lambda(\hat y(J))\le 3\lambda(X_J)+Z+11$, and since $\lambda(\hat y(J))>4Z$ by~\eqref{eq:ch}, we have $\lambda(X_J)>Z-4$ and $\lambda(\hat y(J))\le 5\lambda(X_J)$. As $Z$ is large,
\begin{equation}\label{eq:large}
X_J\ge 2^{\lambda(X_J)-2}>2K_0(k+2)\Phi\lambda(X_J)^3(5\Phi\lambda(X_J))^{c_1}.
\end{equation}

\begin{claim}[content-paid homes]\label{cl:content}
The total charge of the content-paid homes is at most $n$.
\end{claim}
\begin{proof}
Spread $\mathrm{Ch}(J)$ evenly over the vertices of $\mathrm{Reg}(J)$. Each of them receives at most $1/(\Phi\lambda(X_J)^3)$, which is at most $1/\Phi$. By Claim~\ref{cl:mult}, a vertex receives at most $3/8$ from the homes of the first kind. From the homes at each of the at most $\Phi/4$ areas and wraps above it, it receives at most $\sum_{j\ge 1}j^{-3}/\Phi<1.21/\Phi$. In total a vertex receives less than $3/8+1.21/4<1$.
\end{proof}

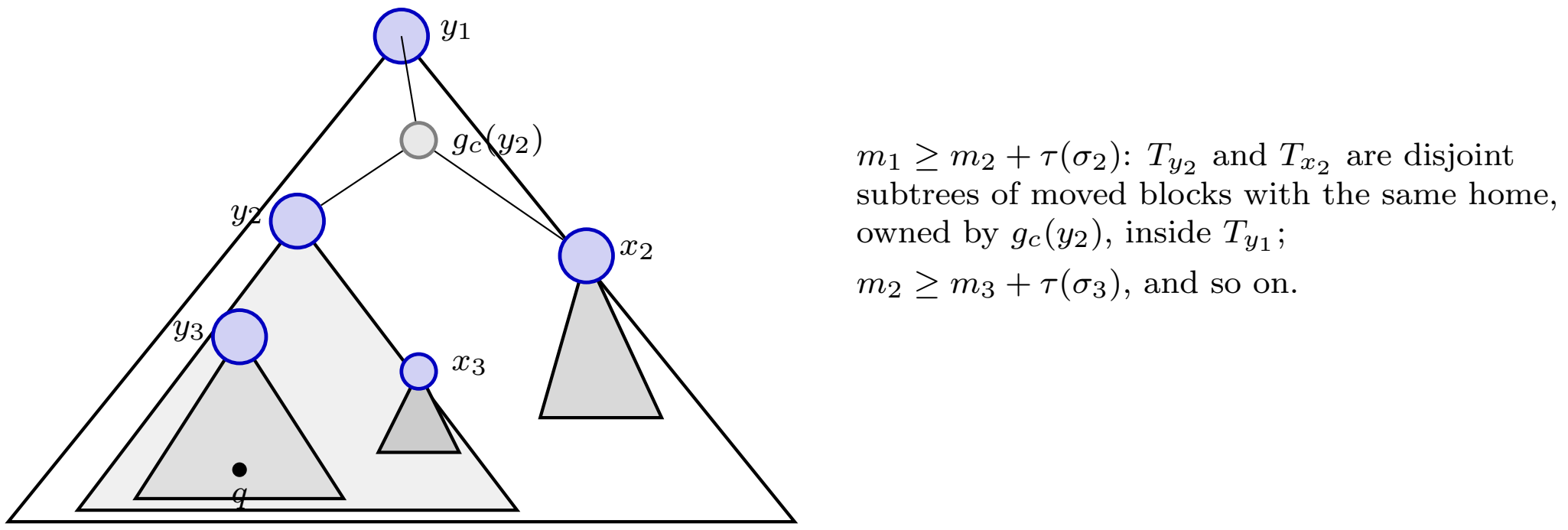
\begin{figure}[t]
\centering
\begin{tikzpicture}[x=1cm,y=1cm]
\draw[thick] (0,0)--(-3.4,-4.2)--(3.4,-4.2)--cycle;
\node[vtx=cB] at (0,0) {};
\node[font=\scriptsize,anchor=west] at (0.2,0.05) {$y_1$};
\node[vtx=gray,minimum size=0.3cm] (g2) at (0.15,-0.9) {};
\node[font=\scriptsize,anchor=west] at (0.3,-0.9) {$g_c(y_2)$};
\draw[thick,fill=gray!12] (-0.9,-1.6)--(-2.8,-4.1)--(1.0,-4.1)--cycle;
\node[vtx=cB] (y2) at (-0.9,-1.6) {};
\node[font=\scriptsize,anchor=east] at (-1.05,-1.55) {$y_2$};
\draw[thick,fill=gray!30] (1.6,-1.9)--(0.95+0.25,-3.3)--(2.25-0.0,-3.3)--cycle;
\node[vtx=cB] (x2) at (1.6,-1.9) {};
\node[font=\scriptsize,anchor=west] at (1.75,-1.85) {$x_2$};
\draw (g2)--(y2); \draw (g2)--(x2); \draw (0,0)--(g2);
\draw[thick,fill=gray!25] (-1.4,-2.6)--(-2.3,-4.0)--(-0.5,-4.0)--cycle;
\node[vtx=cB] at (-1.4,-2.6) {};
\node[font=\scriptsize,anchor=east] at (-1.55,-2.55) {$y_3$};
\draw[thick,fill=gray!40] (0.15,-2.9)--(-0.2,-3.6)--(0.5,-3.6)--cycle;
\node[vtx=cB,minimum size=0.3cm] at (0.15,-2.9) {};
\node[font=\scriptsize,anchor=west] at (0.3,-2.85) {$x_3$};
\fill (-1.4,-3.75) circle (1.8pt); \node[font=\scriptsize,anchor=north] at (-1.4,-3.78) {$q$};
\node[font=\scriptsize,align=left,anchor=west] at (3.8,-1.6) {$m_1\ge m_2+\tau(\sigma_2)$: $T_{y_2}$ and $T_{x_2}$ are disjoint\\ subtrees of moved blocks with the same home,\\ owned by $g_c(y_2)$, inside $T_{y_1}$;\\[3pt] $m_2\ge m_3+\tau(\sigma_3)$, and so on.};
\end{tikzpicture}
\caption{Subtree-paid homes along the path from the root to a vertex $q$. The blocks $B(y_1),B(y_2),\dots$ of one color above $q$ are moved, with subtree-paid homes $J_1,J_2,\dots$. The home of $B(y_{j+1})$ is owned by $g_c(y_{j+1})$, which lies in $T_{y_j}$, and at least one other moved block $B(x_{j+1})$ with the same home has a disjoint subtree of at least $\tau(\sigma_{j+1})$ vertices.}\label{fig:subtree}
\end{figure}

\begin{claim}[subtree-paid homes]\label{cl:subtree}
Every subtree-paid home $J$ has $I(J)\ge 2$, and the total charge of the subtree-paid homes is at most $(k+1)(k+2)K_0(10\Phi)^{c_1}n$.
\end{claim}
\begin{proof}
If $I(J)\le 1$, then by~\eqref{eq:ch} and~\eqref{eq:large}, $\mathrm{Ch}(J)\le 2K_0(k+2)(5\Phi\lambda(X_J))^{c_1}<X_J/(\Phi\lambda(X_J)^3)$, so $J$ is content-paid. Hence $I(J)\ge 2$ for a subtree-paid home. Move the charge of the owner of $J$, if it was assigned to $J$, to one moved block with home $J$. Then every moved block $B(x)$ with home $J$ carries at most $(k+2)K_0(5\Phi\lambda(X_J))^{c_1}$, which we spread evenly over the $\mass(B(x))$ vertices of $T_x\setminus\{x\}$.

Fix a vertex $q$ and a color $c$, and let $y_1,y_2,\dots$ be the ancestors of $q$ of color $c$, from the root down, whose blocks are moved blocks with subtree-paid homes $J_1,J_2,\dots$ (Figure~\ref{fig:subtree}). Let $m_j=\mass(B(y_j))$, $\sigma_j=\sigma(J_j)$ and $Y_j=\tau(\sigma_j)$. Since $B(y_j)$ is heavy and $y_j$ is a proper descendant of the owner of the kept block that $J_j$ belongs to, $Y_j\le m_j<\sigma_j$. The home $J_{j+1}$ is owned by $g_c(y_{j+1})$, which is $y_j$ or a descendant of $y_j$. Hence the at least two moved blocks with home $J_{j+1}$ lie in $T_{y_j}\setminus\{y_j\}$ and have disjoint subtrees, and since each of them has mass at least $Y_{j+1}$,
\[
m_j\ge m_{j+1}+Y_{j+1}.
\]
Together with $m_j\ge Y_j$, this gives $m_j\ge T_j/2$, where $T_j=\sum_{j'\ge j}Y_{j'}$. By~\eqref{eq:large}, $\sigma_j>X_{J_j}\ge 2^{Z-6}$, and since $Z$ is large compared with $\beta\log\beta$, we have $Y_j=\lceil\log\sigma_j\rceil^\beta$ and $\lambda(X_{J_j})\le\lambda(\sigma_j)\le 2Y_j^{1/\beta}$. So $q$ receives through the color $c$ at most
\[
(k+2)K_0(10\Phi)^{c_1}\sum_j\frac{Y_j^{\gamma}}{m_j}\le 2(k+2)K_0(10\Phi)^{c_1}\sum_j\frac{Y_j^{\gamma}}{T_j},\qquad\gamma=\frac{c_1}{\beta}.
\]
Let $B_0=(Z-6)^\beta$, so that $Y_j\ge B_0$ for all $j$, and $B_{b+1}=B_b^2$. Group the indices $j$ into bands $Y_j\in[B_b,B_b^2)$. In band $b$ we have $\log\sigma_j<B_b^{2/\beta}$, and the masses $m_j<\sigma_j$ are distinct, so the band has at most $r_b=2^{B_b^{2/\beta}}$ members, and the $i$-th last of them has $T_j\ge iB_b$. The band therefore contributes at most
\[
\sum_{i=1}^{r_b}\frac{B_b^{2\gamma}}{iB_b}\le B_b^{2\gamma-1}(1+\ln r_b)\le 3B_b^{2\gamma+2/\beta-1}\le 3B_b^{-1/2},
\]
since $2\gamma+2/\beta\le 1/2$ for $\beta\ge 4c_1+4$. Summing over $b$ gives at most $1/2$, as $B_0$ is large. Hence $q$ receives at most $(k+2)K_0(10\Phi)^{c_1}$ through each color, and at most $(k+1)(k+2)K_0(10\Phi)^{c_1}$ in total.
\end{proof}

\end{document}